\documentclass[10pt,conference]{IEEEtran}
\usepackage{bbm}
\IEEEoverridecommandlockouts
\usepackage{cite}
\usepackage{pgfplots}
\pgfplotsset{compat=1.18}
\usetikzlibrary{plotmarks}
\usepackage{amsmath, amssymb, amsfonts}
\usepackage{graphicx}
\usepackage[caption=false,font=footnotesize]{subfig}
\usepackage{caption}
\usepackage{mathtools}

\usepackage{cuted}
\usepackage{tikz}
\usepackage[utf8]{inputenc}
\usepackage[T1]{fontenc}
\usepackage{graphicx}
\usepackage{textcomp}
\usepackage{xcolor}
\usepackage{algorithm}
\usepackage{algpseudocode}
\usepackage{cuted}
\usepackage{bbm}
\usepackage[capitalize]{cleveref}
\usepackage{amsmath, amssymb, amsthm,algorithm}
\usepackage{siunitx}
\usepackage{latexsym}
\usepackage{graphics,epsfig}
\usepackage{amsfonts}
\usepackage{booktabs}
\usepackage{color}
\usepackage{multirow}
\usepackage[justification=centering]{caption}
\usepackage{dsfont}
\usepackage{graphicx}
\usepackage{subcaption} 
\usepackage{adjustbox}
\usepackage{kantlipsum}
\usepackage[caption=false,font=footnotesize]{subfig}
\usepackage{cases}
\usepackage{url}
\usepackage{makecell}
\usepackage{soul}

\usepackage{array}
\usepackage{algpseudocode}
\usepackage{mathtools,lipsum,cuted}

\usepackage[justification=centering]{caption}

\usepackage[caption=false,font=footnotesize]{subfig}
\usepackage[T1]{fontenc}
\usepackage{mwe}
\usepackage{subfig}
\usepackage{pdflscape}

\usepackage{glossaries}
\makeglossaries

\newacronym{tinyml}{TinyML}{Tiny Machine Learning}
\newacronym{iot}{IoT}{Internet of Things}
\newacronym{iotdevice}{IoT device}{IoT device}
\newacronym{ak}{$AIs$}{Intermediate Activations}
\newacronym{si}{SI}{Split-Inference}
\newacronym{mb}{$M_b$}{heartbeat messages}
\newacronym{rk}{$r_k$}{replica of a primary TinyML model segment}
\newacronym{dk}{$s_k$}{primary TinyML model segment}
\newacronym{md}{$M_d$}{IoT device flash memory}
\newacronym{mk}{$m_k$}{size of replica model segment}
\newacronym{e}{$E$}{edge server}
\newacronym{MB}{$M_d^{\mathrm{rep}}$}{total memory occupied by replicas on device $d$}
\newacronym{ota}{$OTA$}{over-the-air}
\newacronym{tb}{$T_b$}{predefined heartbeat period}
\newacronym{e2e}{E2E}{end-to-end}
\newacronym{milp}{MILP}{Mixed-Integer Linear Programming}
\newacronym{mcu}{MCU}{Microcontroller Unit}
\newacronym{ram}{RAM}{random-access Memory}
\newacronym{dnn}{DNN}{Deep Neural Network}
\newacronym{cnn}{CNN}{Convolutional Neural Network}
\newacronym{sram}{SRAM}{Static Random-Access Memory}
\newacronym{ai}{AI}{Artificial Intelligence}

\glsenableentrycount
\makeglossaries

\newlength{\tempheight}
\newlength{\tempwidth}

\newcommand{\rowname}[1]
{\rotatebox{90}{\makebox[\tempheight][c]{#1}}}

\newcommand{\columnname}[1]
{\makebox[\tempwidth][c]{#1}}

\makeatletter
\def\BState{\State\hskip-\ALG@thistlm}
\makeatother

\newcounter{example}[section]

\usepackage[english]{babel}
\usepackage{verbatim}

\usepackage[english]{babel}

\theoremstyle{plain}

\theoremstyle{remark}

\def\BibTeX{{\rm B\kern-.05em{\sc i\kern-.025em b}\kern-.08em
    T\kern-.1667em\lower.7ex\hbox{E}\kern-.125emX}}
\begin{document}

\voffset=0.05in
\textheight=9.28in

\title{Replica-Aware Split Inference for Edge IoT Systems}

\author{\IEEEauthorblockN{Zied Jenhani, Nasrin Razmi, and Admela Jukan}
\IEEEauthorblockA{Technische Universit\"at Braunschweig, Germany;
\{zied.jenhani, n.razmi, a.jukan\}@tu-braunschweig.de}

}

\maketitle

\begin{abstract}

Split-Inference~(SI) addresses the hardware resource limitations of  Internet of Things~(IoT) devices by distributing tiny Machine Learning~(TinyML) model segments across multiple IoT devices. However, a device failure can interrupt the inference pipeline. This paper proposes a replica-aware recovery approach that deploys replicas of TinyML model segments originally hosted
on failed devices onto surviving devices. We formulate the joint selection of replica hosts and communication protocols as a mixed-integer linear programming (MILP) problem that minimizes post-reconfiguration inference latency, accounting for computation and transmission latencies, under flash memory and runtime memory constraints.  
Simulations using MobileNetV2 on heterogeneous IoT networks compare the proposed approach with four baselines. We implement the model in a network of 15 IoT devices with a maximum link-rate reduction of 75\%, and show that MILP approach achieves an average modeled end-to-end inference latency that is 6.8\% lower than that of Computation Greedy and 18.3\% lower than that of Random.
\end{abstract}

\section{Introduction}\label{sec:intro}
\cgls{tinyml} enables on-board inference on resource-constrained
Internet of Things (IoT) devices, reducing dependence on
remote processing \cite{DBLP:journals/corr/abs-2106-07597}. However, larger models may exceed the processing, flash memory, and runtime memory budgets of individual devices. Model optimization and memory-aware inference address these limitations~\cite{9049084,lin2020mcunet}. Complementing these methods, \cgls{si} distributes sequential model segments across \cgls{iot} devices that exchange intermediate activations, combining their memory and processing resources~\cite{jenhani_split_latency}.
The resulting pipeline depends on the availability of all participating devices; a device failure therefore prevents inference from completing. Replica-based approaches improve inference resilience through model redundancy \cite{liu2025frim,guo2024robustdice,wang2024failure,guo2024easter}. However, assigning replicas to surviving IoT devices that already host primary segments while satisfying memory constraints and minimizing computation and transmission latency remains a challenge.

In this paper, we address this challenge with a replica-aware recovery approach that deploys replicas on demand on surviving devices through \cgls{ota} updates, preserving the original partitioning and the placement of surviving primary segments. We formulate the joint selection of replica hosts and communication protocols as a \cgls{milp} problem that minimizes post-reconfiguration inference latency, accounting for computation and transmission latencies, subject to flash memory application-slot and runtime memory capacity constraints. Simulations compare the approach with Random, Computation Greedy, SRAM Worst-Fit, and Flash First-Fit Decreasing across network sizes and failure scenarios.



\section{Related Work}\label{sec:related}
Recent work has studied efficient model partitioning for \cgls{si} across resource-constrained \cgls{iot} devices to support \cgls{tinyml} inference close to sensors~\cite{10454775,hafi2024split}. For example, DeepThings combines fused tile partitioning (FTP) with dynamic task allocation across \cgls{iot} edge devices to reduce execution latency and memory usage~\cite{zhao2018deepthings}. Neuron-level partitioning of convolutional and linear layers enables inference across networked \cglspl{mcu} while reducing per-device peak \cgls{ram} usage~\cite{lu2026splitcnninferencenetworked}. In addition, DISNET combines vertical and horizontal model partitioning to optimize inference latency and energy consumption across \cgls{iot} devices~\cite{samikwa2024disnet}.

These approaches generally assume that all participating devices remain available. Additional recovery mechanisms are therefore needed to sustain distributed inference under device failures. For example, the CONTINUER framework~\cite{9860277} selects recovery mechanisms after edge-node failures based on latency, accuracy, and service downtime requirements. Other resilience approaches include model repartitioning~\cite{9610341}, early-exit mechanisms~\cite{7900006}, and skip connections enabled by gating networks inserted between layers~\cite{wang2018skipnet}. Another approach estimates the predictions of failed devices at the edge using classifier weights and feature vectors from surviving devices, without model replication~\cite{jamshidi2025modifier}.

Beyond these recovery mechanisms, replica-based approaches have also been studied to improve the resilience of distributed \cgls{dnn} execution. The FRIM framework maintains redundant copies of \cgls{dnn} partitions across available nodes and adaptively adjusts the redundancy level to preserve complete inference paths under device failures~\cite{liu2025frim}. A related approach has also been explored for distributed \cgls{dnn} training~\cite{sen2024fault}. Rather than adapting redundancy after deployment, this approach predicts the fault probability of each model partition, determines the required number of replicas accordingly, and assigns these replicas to edge devices before anticipated failures. Other approaches introduce redundancy at a finer granularity or adapt it to specific model architectures. RobustDiCE~\cite{guo2024robustdice} combines partial replication of important neurons with importance-aware partitioning to improve \cgls{cnn} resilience. At the model level, RoCoIn~\cite{wang2024failure} uses knowledge distillation to create compact student models and deploys redundant copies across heterogeneous edge devices. EASTER~\cite{guo2024easter} applies a similar redundancy concept to Transformer models by replicating a selected fraction of important connections while distributing the remaining connections across devices.

Prior work improves resilience through pre-deployed replicas,
redundant sub-models, or replication of neurons and connections.
We address post-failure recovery of vertically partitioned
\cgls{tinyml} inference across heterogeneous \cgls{iot} devices
by deploying replicas on demand through differential
\cgls{ota} updates. Replica hosts and communication protocols
are jointly selected to minimize post-reconfiguration inference
latency under flash memory and runtime memory constraints.
\section{System Model}
\subsection{IoT Network Configuration and Replica-Aware Recovery}
We consider an \cgls{iot} network consisting of an edge server $E$ and a set
$\mathcal{D}=\{1,2,\dots,D\}$ of $D$ geographically static, heterogeneous \cgls{iot}
devices that communicate wirelessly to perform an \cgls{si} task. The fixed distance between any two connected devices
$i,j\in\mathcal{D}$ is denoted by $d_{i,j}$.

The edge server prepares the \cgls{tinyml} model by performing model training, model partitioning, post-training quantization, firmware generation, and \cgls{ota} deployment to \cgls{iot} devices.

It is assumed that the \cgls{si} model has already been partitioned between the \cgls{iot} devices. In this regard, the \cgls{tinyml} model is divided into $K$ vertically
partitioned sequential segments, each of which may contain one or more
layers depending on the selected split points.
These segments are
assigned to the $K$ \cgls{iot} devices, with $K \leq D$. The segment assigned to the $k$-th \cgls{iot} device is denoted by $s_k$ and referred to as its primary \cgls{tinyml} model segment.

 
 In the proposed workflow, the source \cgls{iot} device collects the sensor data and executes its assigned primary \cgls{tinyml} model segment, $s_k$. It then transmits the resulting intermediate activations, $a_k$, obtained from the output of the final layer of segment $s_k$, to the next connected device, which uses $a_k$ as input to execute its assigned primary \cgls{tinyml} model
segment. This process is repeated across the participating \cgls{iot} devices until the destination device produces the final inference result.
 For simplicity, we assume that the source \cgls{iot} device, which hosts the sensor and the first segment $s_1$, does not fail.


\begin{figure}[t]
  \centering
  \includegraphics[width=\columnwidth]{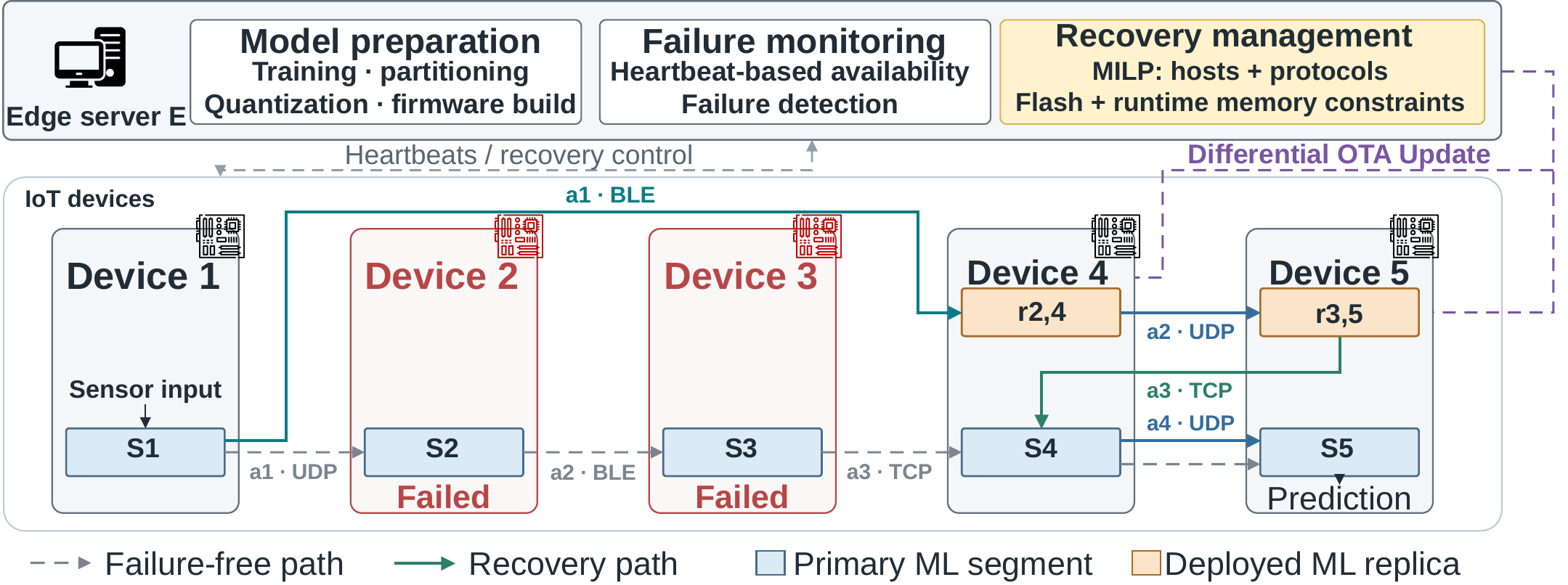}
  \caption{Architecture of the replica-aware split-inference framework.}
  \label{fig:arch1}
\end{figure}
After deployment, the edge server monitors \cgls{iot} devices through heartbeat messages transmitted at a predefined interval $T_b$. An \cgls{iot} device is considered to have failed if no heartbeat message is received within a timeout $T_o \geq T_b$. Upon detecting a failure, the edge server selects a suitable active \cgls{iot} device based on its resource constraints and characteristics, including flash memory, runtime memory, and computational latency, as well as network information such as transmission latency.
Let $r_{k,k'}$ denote the replica of the primary TinyML model segment $s_k$, originally assigned to a failed device $k\in\mathcal F$, that is deployed to a surviving device $k'\in\mathcal S$ for failure recovery, where $\mathcal{F}\subseteq\mathcal{D}$ denotes the set of failed \cgls{iot} devices, and $\mathcal S=\mathcal{D}\setminus\mathcal{F}$ denotes the set of surviving active devices. The replica segment is deployed to the selected device via a differential \cgls{ota} update Fig.~\ref{fig:arch1}. Furthermore, let $\mathcal{R}_{\mathcal F}$ denote the set of all feasible replica-placement configurations for the failed-device set, where each configuration specifies how the \cgls{tinyml} model segments associated with the failed devices are assigned as replicas to the surviving devices in $\mathcal S$. 
{\vspace{-1.5ex}
\subsection{Hardware resources model}}
\paragraph{Flash memory}
Each \cgls{iot} device $k\in\mathcal{D}$ uses a dual-slot \cgls{ota} scheme. Its application-slot capacity $M_k^{\mathrm{OTA}}$ accommodates the fixed flash memory allocation $M_k^{\mathrm{base}}$,
the primary \cgls{tinyml} segment $M_k^{\mathrm{prim}}$, and replicas with a total footprint of $M_k^{\mathrm{rep}}$. The fixed allocation includes the baseline firmware and space unavailable for model storage.  The flash memory footprint of the TinyML
model segment $s_k$, following~\cite{MLSYS2020_0c8abcf1}, is given by
\begin{equation}
m_k =
\sum_{\ell\in\mathcal{L}_k}
\left[
\operatorname{mem}\!\left(w_\ell,Q_\ell\right)
+
M_\ell
\right],
\label{eq:segment_flash}
\end{equation}
where $\mathcal{L}_k$ denotes the set of layers in segment $s_k$,
$w_\ell$ is the weight tensor of layer $\ell$, and $Q_\ell$ denotes its bit precision. The term
$\operatorname{mem}(w_\ell,Q_\ell)$ represents the flash memory footprint of the
quantized weights, while $M_\ell$ accounts for biases, quantization scales,
zero points, and other static parameters associated with layer $\ell$.
Accordingly, $M^{\mathrm{prim}}_{k}=m_k$ for every device hosting a primary
segment, and $M^{\mathrm{prim}}_{k}=0$ otherwise. Since a replica $r_{k,k'}$
is an exact copy of $s_k$, its footprint is also $m_k$.

For a given replica-placement configuration in $\mathcal{R}_{\mathcal{F}}$, let
\begin{equation}
\mathcal{F}_{k'} = \left\{ k\in\mathcal{F} \mid r_{k,k'} \text{ is assigned to device } k' \right\}
\end{equation}
denote the set of failed devices whose primary segments are replicated on the
surviving device $k'\in\mathcal{S}$. The total memory footprint of the replicas
hosted by device $k'$ is then
\begin{equation}
M^{\mathrm{rep}}_{k'} = \sum_{k\in\mathcal{F}_{k'}} m_k,
\qquad \forall k'\in\mathcal{S}.
\end{equation}

The total flash memory usage of each surviving device must remain
within its application partition capacity as
\begin{equation}\label{eq:flash}
M^{\mathrm{base}}_{k'} + M^{\mathrm{prim}}_{k'} + M^{\mathrm{rep}}_{k'}
\leq M^{\mathrm{OTA}}_{k'}, \qquad \forall k'\in\mathcal{S}.
\end{equation}

\paragraph{Runtime memory}
Each \cgls{iot} device is constrained by its memory capacities $M_k^r$, where $r\in\{\mathrm{SRAM},\mathrm{PSRAM}\}$ indexes the
memory pool: \cgls{sram} or
pseudo-static random-access memory (PSRAM). At runtime, this working memory is partitioned between (i) the TinyML execution arena, which maintains intermediate activation tensors, temporary operator scratch buffers, and persistent model states, and (ii) the system and communication overhead. Following the \cgls{sram} decomposition proposed in~\cite{11476185}, we characterize the runtime tensor-arena requirement of the
TinyML segment $s_k$ as
$u_k = u^{\mathrm{act}}_k + u^{\mathrm{scr}}_k + u^{\mathrm{state}}_k$, 
where $u^{\mathrm{act}}_k$ denotes the peak memory occupied by live activation
tensors, $u^{\mathrm{scr}}_k$ denotes the temporary memory required by the
operator kernels, and $u^{\mathrm{state}}_k$ denotes the persistent runtime
state and arena-resident metadata allocated by the inference runtime for $s_k$.
Since a replica $r_{k,k'}$ is an exact copy of $s_k$, it has the same tensor-arena
requirement $u_k$ on its host device $k'$. The activation term is estimated from
the worst-case working-memory requirement during the execution of $s_k$ as
\begin{equation}
u^{\mathrm{act}}_k \approx \max_{\tau}
\sum_{\mathbf{x}\in\mathcal{X}_k(\tau)} \operatorname{size}(\mathbf{x}),
\end{equation}
where $\tau$ denotes the discrete execution step of the layers in $s_k$,
$\mathcal{X}_k(\tau)$ is the set of activation tensors simultaneously alive in
the tensor arena at step $\tau$, and $\operatorname{size}(\mathbf{x})$ is the storage
footprint of tensor $\mathbf{x}$. We extend the segment-level model to each surviving device. For each host considered, $u_k^r$ denotes the segment's
memory requirement in pool $r$ under that host's fixed
arena allocation, including per-segment overhead.
The total memory required by replicas hosted on device
$k'\in\mathcal{S}$ is
\begin{equation}
U_{k'}^{\mathrm{rep},r}
=\sum_{k\in\mathcal{F}_{k'}}u_k^r,
\qquad \forall k'\in\mathcal{S}.
\end{equation}
The total memory usage of device $k'$ is then
\begin{equation}\label{eq:ram}
U_{k'}^r=U_{k'}^{\mathrm{sys},r}
+U_{k'}^{\mathrm{comm},r}
+U_{k'}^{\mathrm{prim},r}
+U_{k'}^{\mathrm{rep},r},
\end{equation}
where $U_{k'}^{\mathrm{sys},r}$ is reserved for system and application tasks,
$U_{k'}^{\mathrm{comm},r}$ is allocated to wireless communication buffers, and
$U_{k'}^{\mathrm{prim},r}$ denotes the memory required by the primary segment of device $k'$, i.e., $U_{k'}^{\mathrm{prim},r}=u_{k'}^r$ if $k'$ hosts a primary segment and $U_{k'}^{\mathrm{prim},r}=0$ otherwise. All quantities refer to pool $r$. Each memory pool must satisfy 
\begin{equation}
U_{k'}^r\le M_{k'}^r,
\quad \forall k'\in\mathcal{S},\
r\in\{\mathrm{SRAM},\mathrm{PSRAM}\}.
\end{equation}

\subsection{Computation latency}

The computation latency of the primary
\cgls{tinyml} model segment $s_k$ executed on its assigned \cgls{iot} device $k$
is estimated relative to a reference model $\Theta$ as
\begin{equation}
T^{\mathrm{comp}}_{k}
=
\frac{v_k}{\phi_{k}}
\left(\frac{\omega_k}{\omega_{\Theta}}\right),
\label{eq:comp_latency_prim}
\end{equation}
where $v_k$ is the data volume of the input activation $a_{k-1}$ of segment
$s_k$, $\phi_{k}$ is the computing rate of device $k$, and
$\omega_k$ and $\omega_\Theta$ denote the numbers of computational operators in segment $s_k$ and the reference model $\Theta$, respectively.

When device $k\in\mathcal{F}$ fails, its segment runs as the replica
$r_{k,k'}$ on the surviving host device $k'\in\mathcal{S}$. Since $r_{k,k'}$ is
an exact copy of $s_k$, it has the same input volume $v_k$ and operator count $\omega_k$, while its execution speed is determined by the host device.
Accordingly, the computation latency of $r_{k,k'}$ is
\begin{equation}
T^{\mathrm{comp}}_{k,k'}
=
\frac{v_k}{\phi_{k'}}
\left(\frac{\omega_k}{\omega_{\Theta}}\right),
\qquad \forall k\in\mathcal{F},\ k'\in\mathcal{S}.
\label{eq:comp_latency_rep}
\end{equation}
With vertical, layer-wise partitioning, each segment processes the full input feature map. Thus, the data volume of the input
activation $a_{k-1}$ of segment $s_k$ is
$v_k = H_k \, W_k \, C_k \, b,$
where $H_k$, $W_k$, and $C_k$ denote the height, width, and channel
count of $a_{k-1}$, and $b$ is the number of bytes per tensor
element, with $b = 1$ under INT8 quantization. The computing rate of the $k$-th \cgls{iot} device, expressed in
bytes per second, is
$\phi_{k} = \frac{f_{k}}{c_{k}}$,
where $f_{k}$ and $c_{k}$ denote the clock frequency in cycles per
second and the number of clock cycles required to process
one byte.

\subsection{Transmission latency}

The transmission latency represents the time required to transfer intermediate activations
$a_k$ from segment $s_k$ to $s_{k+1}$. In the failure-free case, $a_k$ is
sent from device $k$ to device $k+1$. If device $k$ fails, $s_k$ is executed
as the replica $r_{k,k'}$ on the surviving device $k'$, which then becomes the
sender; similarly, if device $k+1$ fails, $s_{k+1}$ is executed as the replica
$r_{k+1,k''}$ on the surviving device $k''$, which then becomes the receiver.
Let $\mathcal{P}$ denote the set of considered communication
protocols. For each pair of distinct surviving devices
$i,j \in \mathcal{S}$, we define
$\mathcal{P}_{i,j} \subseteq \mathcal{P}$ as the set of
protocols supported by both devices and usable over the
direct link from $i$ to $j$. These sets are assumed known
to the edge server and fixed during the evaluated inference.
An inter-device activation transfer can use only a protocol
$p \in \mathcal{P}_{i,j}$. If
$\mathcal{P}_{i,j} = \varnothing$, direct communication
between the two devices is unavailable. When consecutive
segments execute on the same device, their activation is
passed locally with zero communication latency.\\
Assuming a direct wireless link using an available
protocol $p$, the transmission latency is
\begin{equation}
T_k^{\mathrm{tr},p} =
\begin{cases}
0, & i=j, \\[1mm]
\displaystyle \frac{v_{k+1}}{B_{i,j}^{p}},
& i\ne j,\ p\in\mathcal{P}_{i,j},
\label{eq:tx_latency}
\end{cases}
\end{equation}
where $v_{k+1}$ denotes the size of the intermediate activation $a_k$ and $B^{p}_{i,j}$ denotes the effective
throughput (bytes/s) between the corresponding transmitter $i$ and receiver $j$ under
protocol $p$. Here, $i=k$ if device $k$ is active, and $i=k'$ if device $k$ has failed and its segment is executed on the surviving replica device $k'$. Similarly, $j=k+1$ if device $k+1$ is operational, and $j=k''$ if the segment is executed on the surviving replica device $j=k''$.
For $i=j$, no wireless transfer is required.\\
{\vspace{-2ex}
\section{Problem Formulation}}
The objective of the Replica-Aware recovery approach is to assign each failed segment to a surviving device hosting its replica, minimizing
\cgls{e2e} inference latency while satisfying the flash memory and runtime memory constraints.

Let $\mathcal{F}^{K}=\mathcal{F}\cap\{1,\dots,K\}$ denote the set of failed
devices that hosted a primary segment. The replica placement is described by
the binary decision variables $x_{k,k'}\in\{0,1\}$, where $x_{k,k'}=1$ if the
replica $r_{k,k'}$ of segment $s_k$, $k\in\mathcal{F}^{K}$, is deployed on the
surviving device $k'\in\mathcal{S}$, and $x_{k,k'}=0$ otherwise. The placement
configuration is denoted by $\mathbf{x}=[x_{k,k'}]$, and the set of replicas
hosted by device $k'$ is
$\mathcal{F}_{k'}(\mathbf{x})=\{k\in\mathcal{F}^{K}\mid x_{k,k'}=1\}$.
Let $p=(p_1,\ldots,p_{K-1})$ denote the
communication choices at the segment boundaries, where
$p_k$ is the protocol used to transfer activation $a_k$
between distinct hosts. For co-located segments, we use
$p_k=\mathrm{local}$ to indicate that no wireless transfer
is required.\\
The \cgls{e2e} inference latency of the reconfigured pipeline is
\begin{equation}
\begin{aligned}
T^{\mathrm{e2e}}(\mathbf{x,p})
={}& \sum_{k\in\{1,\dots,K\}\setminus\mathcal{F}^{K}} T^{\mathrm{comp}}_{k}
+ \sum_{k\in\mathcal{F}^{K}}\sum_{k'\in\mathcal{S}} x_{k,k'}\,T^{\mathrm{comp}}_{k,k'} \\
&+ \sum_{k=1}^{K-1} T^{\mathrm{tr},p_{k}}_{k}(\mathbf{x}),
\end{aligned}
\label{eq:e2e_latency}
\end{equation}
where the first term is the computation latency of the primary segments on the
active devices, the second term is the computation latency of the replicas on
their selected host devices, and the third term is the transmission latency of
the intermediate activations along the reconfigured pipeline, as defined
in~\eqref{eq:tx_latency}. For each inter-device transfer, $p_k$ must belong to $\mathcal{P}_{i,j}$, where $i$ and $j$ are the actual hosts determined by placement $x$. For co-located
segments, the transmission term is zero.
For a given placement $x$, let $\mathcal{A}(x)$ denote
the set of admissible communication-choice vectors compatible with the replica-placement. The replica-placement and protocol-selection problem is formulated as
\begin{subequations}\label{eq:problem}
\begin{align}
\min_{\mathbf{x,p}}\;\; & T^{\mathrm{e2e}}(\mathbf{x,p}) \label{eq:obj}\\
\text{s.t.}\;\; 
& \textstyle\sum_{k'\in\mathcal{S}} x_{k,k'} = 1,\;\; \forall k\in\mathcal{F}^{K}, \label{eq:c_assign}\\
& M^{\mathrm{base}}_{k'} + M^{\mathrm{prim}}_{k'}
  + \textstyle\sum_{k\in\mathcal{F}^{K}} x_{k,k'}\,m_k
  \nonumber\\
& \quad \le M^{\mathrm{OTA}}_{k'},\;\;
  \forall k'\in\mathcal{S}, \label{eq:c_flash}\\
& U^{\mathrm{sys},r}_{k'} + U^{\mathrm{comm},r}_{k'}
  + U^{\mathrm{prim},r}_{k'}
  + \textstyle\sum_{k\in\mathcal{F}^{K}} x_{k,k'}\,u_k^r \nonumber\\
& \quad \le M^r_{k'},\;\; \forall k'\in\mathcal{S},\;
  r\in\{\mathrm{SRAM},\mathrm{PSRAM}\}, \label{eq:c_sram}\\
& \mathbf{x}\in\{0,1\}^{|\mathcal{F}^{K}|\times|\mathcal{S}|},\\
& p \in\mathcal{A}(x),
\label{eq:c_binary}
\end{align}
\end{subequations}
where constraint~\eqref{eq:c_assign} ensures that each failed segment is replicated
on exactly one surviving device. Constraints~\eqref{eq:c_flash}
and~\eqref{eq:c_sram} guarantee that the primary segment and the hosted
replicas of each device fit within its flash memory and
runtime memory capacity, respectively. The feasible set $\mathcal{R}_{\mathcal{F}}$ consists of placements satisfying (13b)--(13e) with at least one compatible protocol assignment. An optimal solution satisfies
\begin{equation}
(x^\star,p^\star)
\in
\arg\min_{\substack{
x\in\mathcal{R}_{\mathcal{F}}\\
\boldsymbol{p}\in\mathcal{A}(x)}}
T^{\mathrm{e2e}}(x,p).
\end{equation}

\section{Solution}\label{sec:solution}

The problem defined in \cref{eq:problem} can be reformulated as an \cgls{milp} by linearizing the placement-dependent communication terms. When both consecutive segments $s_k$ and $s_{k+1}$ require replicas, their link latency depends jointly on the selected hosts, introducing bilinear terms into the transmission latency in \cref{eq:obj}.
To consider all cases of \cref{eq:tx_latency} uniformly, we define for each segment
$s_k$ the candidate-host set
\begin{equation}
\mathcal{H}_k =
\begin{cases}
\{k\}, & k \notin \mathcal{F}^K,\\
\mathcal{S}, & k \in \mathcal{F}^K,
\end{cases}
\end{equation}
and, for $i \in \mathcal{H}_k$, the host indicator
\begin{equation}
h_{k,i} =
\begin{cases}
1, & k \notin \mathcal{F}^K,\ i = k,\\
x_{k,i}, & k \in \mathcal{F}^K,
\end{cases}
\end{equation}
which is either a constant or an existing placement variable and thus
introduces no new decisions.

For each segment boundary $k=1,\ldots,K-1$ and distinct
candidate hosts $i\in\mathcal{H}_k$, $j\in\mathcal{H}_{k+1}$,
we introduce $z_{k,i,j,p}\in\{0,1\}$ for each available
protocol $p\in\mathcal{P}_{i,j}$. This variable equals one
when activation $a_k$ is transferred from device $i$ to
device $j$ using protocol $p$.

The transmission latency of activation $a_k$ then becomes
\[
T_k^{\mathrm{tr},p_k}(x)=
\sum_{i\in\mathcal{H}_k}
\sum_{\substack{j\in\mathcal{H}_{k+1}\\j\ne i}}
\sum_{p\in\mathcal{P}_{i,j}}
z_{k,i,j,p}\frac{v_{k+1}}{B_{i,j}^{p}},
\]
with zero communication latency for co-located hosts
($i=j$), consistent with \cref{eq:tx_latency}.

The products $h_{k,i} h_{k+1,j}$ are bilinear only when both consecutive
segments belong to $\mathcal{F}^K$; otherwise one factor is constant and
the term is already linear. For each bilinear product we apply the standard
linearization of Glover~\cite{glover1975improved}, introducing an auxiliary
variable $y_{k,i,j} \in [0,1]$ with
\begin{equation}
y_{k,i,j} \le h_{k,i}, \quad
y_{k,i,j} \le h_{k+1,j}, \quad
y_{k,i,j} \ge h_{k,i} + h_{k+1,j} - 1,
\end{equation}
which enforces $y_{k,i,j} = h_{k,i} h_{k+1,j}$ at any feasible binary
point. For the remaining candidate-host pairs, where at least
one host indicator is constant, we define
$y_{k,i,j}=h_{k,i}h_{k+1,j}$ directly; this relation is
already linear. Thus, $y_{k,i,j}$ represents the selected
host pair at every segment boundary.

Protocol selection is linked to the host-pair selection by
$
\sum_{p\in\mathcal{P}_{i,j}}z_{k,i,j,p}
=
y_{k,i,j}$, 
for all $k=1,\ldots,K-1$, $i\in\mathcal{H}_k$,
$j\in\mathcal{H}_{k+1}$ with $i\ne j$.
When $y_{k,i,j}=1$, exactly one available protocol is
selected. When $y_{k,i,j}=0$, all corresponding
protocol-selection variables are zero.

These constraints encode $p\in\mathcal{A}(x)$.
For an inter-device transfer, the selected protocol $p_k$
is recovered from $z_{k,i,j,p}=1$; for a local transfer,
$p_k=\mathrm{local}$. Since the transfer coefficients
$v_{k+1}/B_{i,j}^{p}$ are known constants, the resulting
objective and constraints are linear in $x$, $y$, and $z$.

\section{Performance Evaluation}\label{sec:PERFORMANCE EVALUATION}

\subsection{Implementation and Device Parameters}
We ran simulations on an Intel Core i9-14900 workstation
with 64~GB of RAM. Memory capacities and OTA slot sizes
for \cgls{iot} devices follow hardware specifications~\cite{espressif_esp32s3,arduino_nano33,raspberry_pico2w,stm32_bu585} and partition layouts~\cite{espressif_partitions,raspberry_ota,stm32_tfm,mcuboot_design} summarized in Table~\ref{tab:device_parameters}.
For ESP32-S3 devices, the simulated fixed flash memory  allocation leaves $1.3\,\mathrm{MiB}$ for primary and replica
\cgls{tinyml} segments. Compatible differential-OTA
software is assumed~\cite{espressif_delta_ota}.
Segment storage and intermediate activation sizes follow
the model partitions. Tensor arenas~\cite{tflm_memory} use PSRAM on ESP32-S3
and internal SRAM on Arduino, Pico, and STM32 devices,
with an assumed $2\,\mathrm{KiB}$ SRAM overhead per hosted segment.
Latency models use assumed device computing rates for
computation, and activation sizes and effective link rates
referenced to ESP32-S3 benchmarks for
transmission~\cite{jenhani_split_latency}. The simulation scripts and input datasets used in the evaluation
are available in the accompanying GitHub repository 
\footnote{https://github.com/Zied-Jenhani/Replica-Aware-Split-Inference-for-Edge-IoT-Systems}.
\begin{table}[t]
\centering
\caption{Device memory and OTA slot sizes (KiB)}
\label{tab:device_parameters}
\resizebox{\columnwidth}{!}{%
\begin{tabular}{lrrrrr}
\hline
\textbf{Device} &
\textbf{SRAM} &
\textbf{PSRAM} &
\textbf{Internal Flash} &
\textbf{External Flash} &
\textbf{OTA slot} \\
\hline
ESP32-S3-WROOM-1-N16R8
  & 512 & 8192 & -- & 16384 & 7680 \\
Arduino Nano 33 BLE Sense Rev2
  & 256 & -- & 1024 & -- & 448 \\
Raspberry Pi Pico 2 W
  & 520 & -- & -- & 4096 & 1744 \\
STM32 B-U585I-IOT02A
  & 786 & 8192 & 2048 & 65536 & 832 \\
\hline
\end{tabular}}
\end{table}
\subsection{Network and Model Configuration}

We evaluate networks comprising $D \in \{5,10,15\}$ distributed IoT
devices. The device counts, ordered as ESP32-S3 / Arduino / Pico / STM32, are $(2,1,1,1)$
for $D=5$, $(3,3,2,2)$ for $D=10$, and $(4,4,4,3)$
for $D=15$. Intermediate activations are exchanged directly using Wi-Fi (TCP/UDP),
BLE, or ESP-NOW, with ESP-NOW restricted to ESP32-S3 pairs.
Differential \cgls{ota} updates use the server's Wi-Fi access
point for Wi-Fi-capable devices and BLE for the Arduino
Nano 33 BLE Sense. We assume that each device is connected to the server through a wired monitoring interface for periodic status reports and recovery control.  This infrastructure is assumed to remain available and is excluded from $D$ and the inference-device resource budgets. The workload is an ImageNet-pretrained MobileNetV2 model with $\alpha=0.35$, input dimensions of $224\times224\times3$,
and INT8 quantization. We set $K=D$ and keep the initial partitioning fixed
throughout the evaluation.

\subsection{Failure Scenarios and Baselines}

For each network size, we evaluate all $\binom{D-1}{F}$ failure combinations for $F=|\mathcal{F}|\in\{1,\ldots,\min(5,D-1)\}$. The data-acquisition device is excluded from the failure set,
and the initial deployment remains fixed across all failure
combinations. 
Four baselines preserve surviving primaries and enforce flash memory and runtime memory constraints.
{Random} places failed segments in random order on uniformly selected eligible surviving devices; results aggregate three seeds per scenario. {Computation Greedy} and {SRAM Worst-Fit} place segments in pipeline order, respectively minimizing computation latency and maximizing the remaining SRAM fraction after placement.
{Flash First-Fit Decreasing (FFD)} processes segments by decreasing flash memory demand, selecting the first eligible device in ascending device-ID order. Each feasible heuristic placement uses the highest-throughput compatible protocol for transfers between distinct hosts. All methods share input parameters and feasibility checks.
\subsection{Evaluation Metrics and Results}
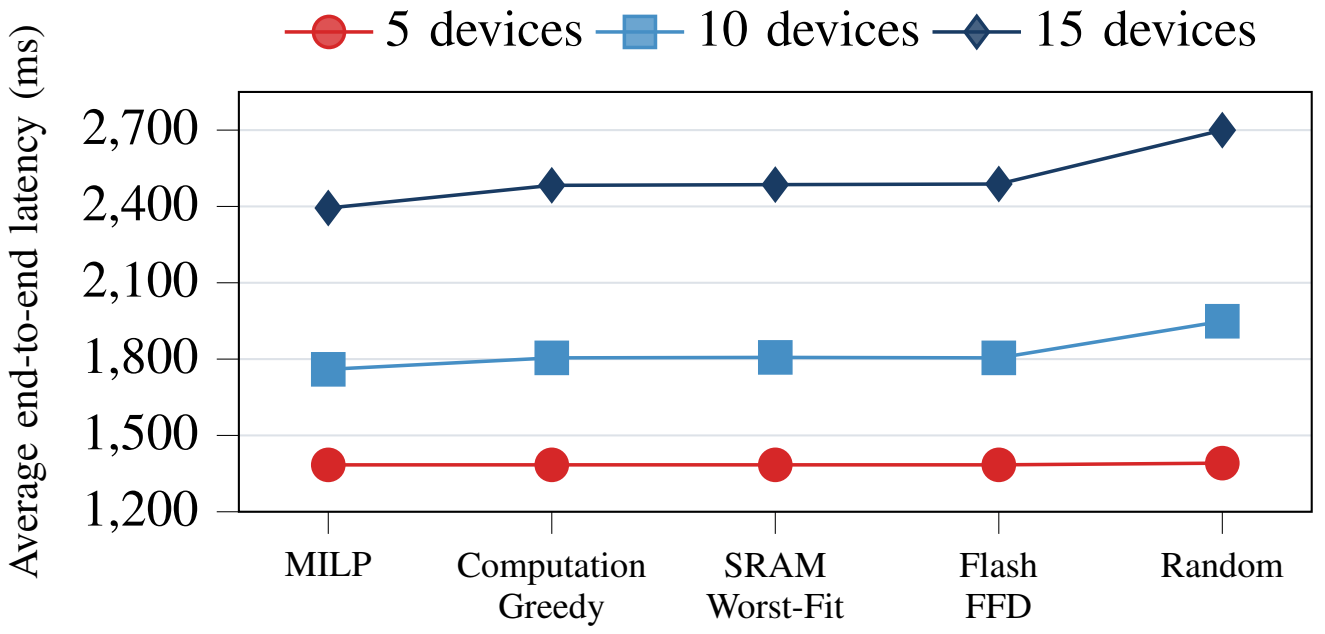
\begin{figure}[t]
  \centering
  \resizebox{\columnwidth}{!}{%
\begin{tikzpicture}

\definecolor{crimson2143940}{RGB}{214,39,40}
\definecolor{darkgray176}{RGB}{176,176,176}
\definecolor{darkslategray255999}{RGB}{25,59,99}
\definecolor{gainsboro221226232}{RGB}{221,226,232}
\definecolor{steelblue70143197}{RGB}{70,143,197}

\begin{axis}[
width=\columnwidth,
height=0.50\columnwidth,
legend cell align={left},
legend columns=3,
legend cell align={left},
legend columns=3,
legend style={
  fill opacity=0.8,
  draw opacity=1,
  text opacity=1,
  at={(0.5,1.04)},
  anchor=south,
  draw=none
},
tick align=outside,
tick pos=left,
x grid style={darkgray176},
xmin=-0.4, xmax=4.4,
xtick style={color=black},
xtick={0,1,2,3,4},
xticklabels={
  {MILP},
  {Computation\\Greedy},
  {SRAM\\Worst-Fit},
  {Flash\\FFD},
  {Random}
},
xticklabel style={align=center,font=\scriptsize},
y grid style={gainsboro221226232},
ylabel={Average end-to-end latency (ms)},
ylabel style={font=\footnotesize},
ymajorgrids,
clip=false,
ymin=1200, ymax=2850,
ytick style={color=black},
ytick={900,1200,1500,1800,2100,2400,2700,3000},
yticklabels={900,{1,200},{1,500},{1,800},{2,100},{2,400},{2,700},{3,000}}
]
\addplot [line width=0.68pt, crimson2143940, mark=*, mark size=3, mark options={solid}]
table {%
0 1384.28494627028
1 1384.39681772896
2 1384.39681772896
3 1384.39681772896
4 1391.06938320021
};
\addlegendentry{5 devices}

\addplot [line width=0.68pt, steelblue70143197, mark=square*, mark size=3, mark options={solid}]
table {%
0 1759.80956075422
1 1804.8492369437
2 1806.69119222029
3 1804.8492369437
4 1948.74893826793
};
\addlegendentry{10 devices}

\addplot [line width=0.68pt, darkslategray255999, mark=diamond*, mark size=3, mark options={solid}]
table {%
0 2393.78054666621
1 2482.92754817554
2 2485.673512757
3 2488.13702677168
4 2699.77618467299
};
\addlegendentry{15 devices}

\addplot [draw=none, draw=white, fill=crimson2143940, forget plot, mark=asterisk]
table{%
x  y
3.06161699786838e-17 0.5
-0.112256990838573 0.154508492636711
-0.475528258147577 0.154508497187474
-0.181635626651597 -0.0590169926367106
-0.293892626146237 -0.404508497187474
-3.50830079462338e-17 -0.190983
0.293892626146236 -0.404508497187474
0.181635626651597 -0.0590169926367106
0.475528258147577 0.154508497187474
0.112256990838573 0.154508492636711
1.53080849893419e-16 0.5
};
\addplot [draw=none, draw=white, fill=steelblue70143197, forget plot, mark=asterisk]
table{%
x  y
3.06161699786838e-17 0.5
-0.112256990838573 0.154508492636711
-0.475528258147577 0.154508497187474
-0.181635626651597 -0.0590169926367106
-0.293892626146237 -0.404508497187474
-3.50830079462338e-17 -0.190983
0.293892626146236 -0.404508497187474
0.181635626651597 -0.0590169926367106
0.475528258147577 0.154508497187474
0.112256990838573 0.154508492636711
1.53080849893419e-16 0.5
};
\addplot [draw=none, draw=white, fill=darkslategray255999, forget plot, mark=asterisk]
table{%
x  y
3.06161699786838e-17 0.5
-0.112256990838573 0.154508492636711
-0.475528258147577 0.154508497187474
-0.181635626651597 -0.0590169926367106
-0.293892626146237 -0.404508497187474
-3.50830079462338e-17 -0.190983
0.293892626146236 -0.404508497187474
0.181635626651597 -0.0590169926367106
0.475528258147577 0.154508497187474
0.112256990838573 0.154508492636711
1.53080849893419e-16 0.5
};
\end{axis}

\end{tikzpicture}%
  }
  \par\vspace{-13mm}
  \caption{Average E2E inference latency at the original link rates.}
  \label{fig:e2elatency}
\end{figure}

\begin{figure}[t]
  \centering
  \resizebox{\columnwidth}{!}{%
\begingroup
\definecolor{siMILP}{HTML}{193B63}
\definecolor{siGreedy}{HTML}{18836C}
\definecolor{siSRAM}{HTML}{A45C91}
\definecolor{siFlash}{HTML}{468FC5}
\definecolor{siRandom}{HTML}{D88700}
\begin{tikzpicture}
\begin{axis}[
  width=\columnwidth,
  height=0.58\columnwidth,
  scale only axis=false,
  font=\fontsize{7}{8}\selectfont,
  axis lines=box,tick pos=left,
  axis line style={black!75},
  tick align=outside,
  tick style={black!75},
  tick label style={font=\fontsize{6}{7}\selectfont},
  xmin=-7, xmax=108,
  ymin=2000, ymax=4500,
  xtick={0,25,50,75},
  xticklabels={0,25,50,75},
  scaled x ticks=false,
  ytick={2000,2500,3000,3500,4000,4500},
  yticklabels={{2,000},{2,500},{3,000},{3,500},{4,000},{4,500}},
  xlabel={Maximum link-rate reduction (\%)},
  ylabel={Average end-to-end latency (ms)},
  xlabel style={font=\fontsize{7}{8}\selectfont},
  ylabel style={font=\fontsize{7}{8}\selectfont},
  ymajorgrids=true,
  grid style={gray!22,line width=0.3pt},
  clip=false,
  legend columns=3,
  legend cell align=left,
  legend style={
    at={(0.5,1.06)}, anchor=south,
    draw=none, fill=none,
    font=\fontsize{6}{7}\selectfont,
    inner sep=0pt,
    /tikz/every even column/.append style={column sep=3pt}
  },
  legend image post style={line width=0.8pt,mark size=1.7pt},
]
\addplot[
  color=siMILP, solid, line width=0.9pt,
  mark=star, mark size=2.4pt
] coordinates {
  (0,2393.7805) (25,2580.7516)
  (50,2855.4098) (75,3361.9222)
};
\addlegendentry{MILP}

\addplot[
  color=siGreedy,dashed,line width=0.75pt,
  mark=o,mark size=1.9pt,mark options={solid,fill=white}
] coordinates {
  (0,2482.9275) (25,2700.6864) (50,3023.5176) (75,3607.9355)
};
\addlegendentry{Computation Greedy}

\addplot[
  color=siSRAM,dashdotted,line width=0.75pt,
  mark=square,mark size=1.9pt,mark options={solid,fill=white}
] coordinates {
  (0,2485.6735) (25,2712.1796) (50,3056.2140) (75,3706.3727)
};
\addlegendentry{SRAM Worst-Fit}

\addplot[
  color=siFlash,densely dotted,line width=0.8pt,
  mark=triangle,mark size=2pt,mark options={solid,fill=white}
] coordinates {
  (0,2488.1370) (25,2708.5436) (50,3037.3177) (75,3638.4639)
};
\addlegendentry{Flash FFD}

\addplot[
  color=siRandom, solid, line width=0.8pt,
  mark=diamond*, mark size=2pt
] coordinates {
  (0,2699.7762) (25,2960.5505)
  (50,3358.5973) (75,4114.6592)
};
\addlegendentry{Random}

\draw[black!45,densely dotted,line width=0.4pt]
  (axis cs:76,3361.9222)--(axis cs:78.5,3361.9222);
\draw[black!45,densely dotted,line width=0.4pt]
  (axis cs:76,3607.9355)--(axis cs:78.5,3607.9355);
\draw[black!60,line width=0.45pt]
  (axis cs:78.5,3361.9222)--(axis cs:78.5,3607.9355);
\draw[black!60,line width=0.45pt]
  (axis cs:77.3,3361.9222)--(axis cs:79.7,3361.9222);
\draw[black!60,line width=0.45pt]
  (axis cs:77.3,3607.9355)--(axis cs:79.7,3607.9355);
\node[
  anchor=west,align=left,
  font=\fontsize{6}{7}\selectfont,
  text=black!85,inner sep=1pt
] at (axis cs:81,3484.92885) {Greedy vs MILP\\\textbf{246.02 ms}};
\end{axis}
\end{tikzpicture}
\endgroup%
  }
  \caption{Average E2E inference latency under link-rate reductions for 15 devices.}
  \label{fig:linkrate}
\end{figure}
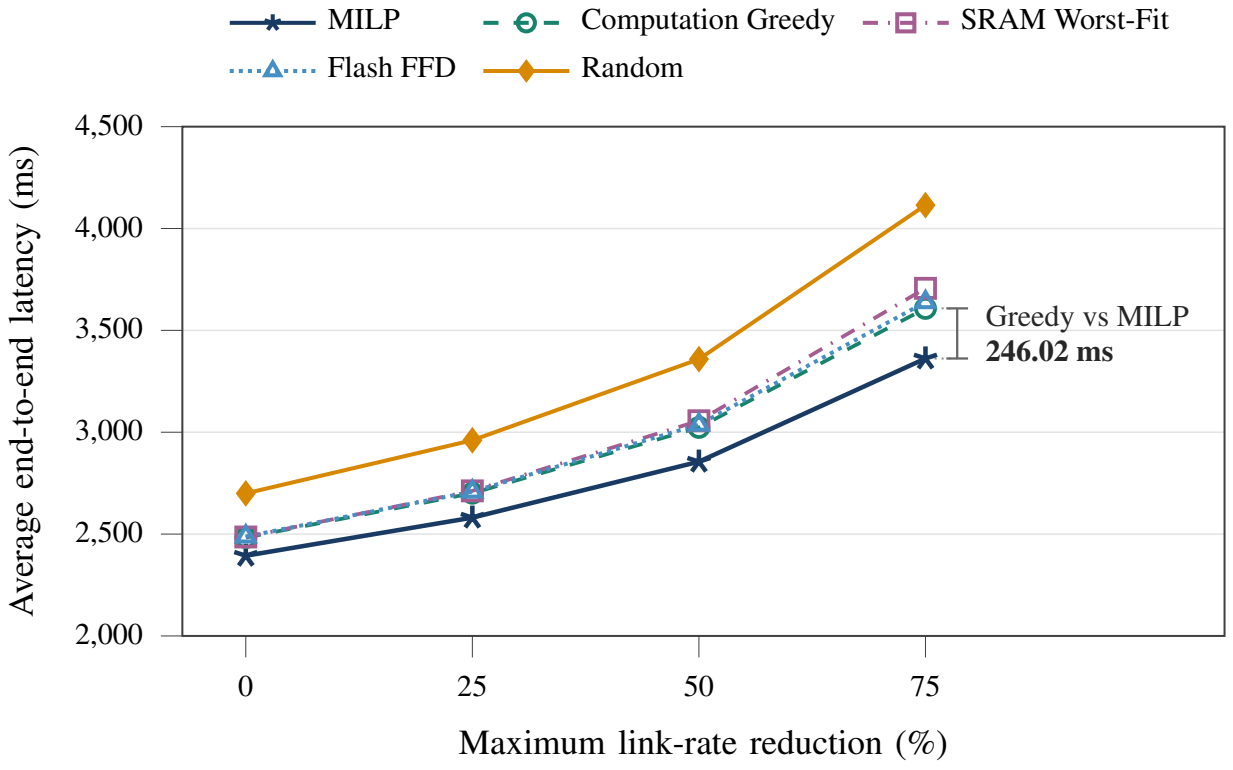

\begin{figure}[t]
  \centering
  \includegraphics[width=\columnwidth]{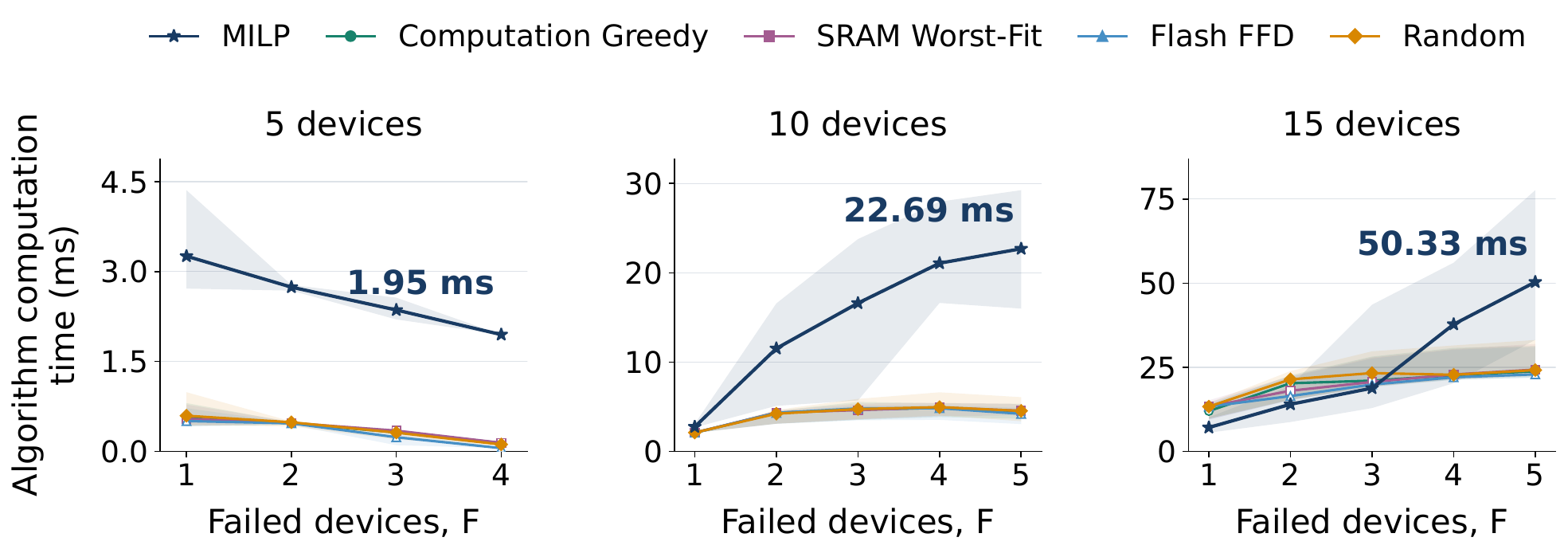}
\caption{Median algorithm computation time versus failure
count, with interquartile bands.}
  \label{fig:computationtime}
\end{figure}

We evaluate three metrics: E2E inference latency \(T^{\mathrm{e2e}}(\mathbf{x,p})\), the algorithm computation time, and recovery success rate, defined as the percentage of attempts that find a valid recovery plan. Latency is evaluated at the original link rates and under
link-specific reductions ranging from zero to a maximum
of 25\%, 50\%, or 75\%. Results are averaged over three
link-rate configurations per reduction level. The average latency is computed only for failure scenarios that every method successfully recovers.
For the \cgls{milp}, computation time includes problem construction, optimization, and verification; for the heuristics, it includes replica placement, protocol selection,
and feasibility verification. We report median workstation execution times with interquartile bands over all completed attempts, including unsuccessful attempts.
Fig.~\ref{fig:e2elatency} shows that the \cgls{milp} achieves the lowest average E2E latency across all evaluated network sizes. Results nearly coincide with the deterministic
heuristics for five devices, with larger differences for ten and fifteen devices. For fifteen devices, the \cgls{milp} achieves \SI{2393.78}{\milli\second}, compared with \SI{2482.93}{\milli\second} for Computation Greedy and \SI{2699.78}{\milli\second} for Random. The \cgls{milp} considers computation and transmission latency jointly when selecting replica hosts. The evaluated heuristics instead use random, computation-based, or memory-based placement rules, followed by selection of the fastest compatible protocols.
Fig.~\ref{fig:linkrate} shows that average E2E latency increases for all methods as link rates decrease. For fifteen devices at a maximum link-rate reduction of 75\%, latency reaches
\SI{3361.92}{\milli\second} for the \cgls{milp}, \SI{3607.94}{\milli\second} for Computation Greedy, and \SI{4114.66}{\milli\second} for Random. Lower rates increase transmission latency, making communication costs more influential in replica placement.
Fig.~\ref{fig:computationtime} shows increasing median \cgls{milp} computation time with failure count for ten and fifteen devices, reaching \SI{22.69}{\milli\second} and
\SI{50.33}{\milli\second}, respectively, at five failures, where the evaluated heuristics are faster. For five devices with four failures, recovery is infeasible under the memory constraints, with
a median \cgls{milp} computation time of \SI{1.95}{\milli\second}. At the original link rates, the \cgls{milp} and deterministic
heuristics achieve identical recovery success rates:
73.3\%, 94.2\%, and 100\% for $D=5,10,15$, respectively.

\iffalse
\newpage
{\color{orange}$d_k^{\mathrm{prim}}\in\mathcal{D}$ denote the fixed primary device
of segment $k$.}
The objective is to minimize the reconfigured end-to-end inference latency
$T^{F}_{\mathrm{E2E}}$ of~\eqref{eq:e2e_post_failure}, subject to the per-device flash memory,
SRAM, and energy budgets of~\eqref{eq:flash},~\eqref{eq:ram},
and~\eqref{eq:energy}. We can model this replica-placement problem as a
\cgls{milp}.

The vector $\mathbf{x}$ realizes the feasible mapping $r\in R(F)$
of~\eqref{eq:argmin}; $\mathbf{y}$ selects the device pair and protocol for each
transfer. All coefficients are known to the edge server at reconfiguration
time.
{\color{orange}
\begin{subequations}\label{eq:p1}
\begin{align}
\textbf{(P1)}\quad
\min_{\bm{x},\bm{y},\bm{z}}\;\;
&
\sum_{k\in\mathcal{S}}
\sum_{d\in\mathcal{D}_s}
x_{k,d}T_{k,d}^{\mathrm{comp}}
\nonumber\\
&+
\sum_{k=1}^{K}
\sum_{i,j\in\mathcal{D}_s}
\sum_{p\in\widetilde{\mathcal{P}}_{i,j}}
y_{k,i,j}^{p}T_{k,i,j}^{\mathrm{rep},p}
\label{eq:objective}\\
\text{s.t.}\quad
&
\sum_{d\in\mathcal{D}_s}x_{k,d}=1,
\quad
\forall k\in\mathcal{S}
\label{eq:assign}\\
&
x_{k,d_k^{\mathrm{prim}}}=1,
\quad
\forall k\in\mathcal{S}\setminus\mathcal{S}_{\mathcal F}
\label{eq:fixed_primary}\\
&
\sum_{c\in\mathcal{C}_d^{\mathcal F}}
z_{d,c}=1,
\quad
\forall d\in\mathcal{D}_s
\label{eq:one_configuration}\\
&
x_{k,d}
=
\sum_{\substack{
c\in\mathcal{C}_d^{\mathcal F}\\
k\in c
}}
z_{d,c},
\quad
\forall k\in\mathcal{S}_{\mathcal F},
\ \forall d\in\mathcal{D}_s
\label{eq:configuration_placement_link}\\
&
M_d^{\mathrm{prim}}
+
\sum_{k\in\mathcal{S}_{\mathcal F}}
m_kx_{k,d}
\leq M_d,
\quad
\forall d\in\mathcal{D}_s
\label{eq:c_flash}\\
&
SR_d^{\mathrm{sys}}
+
SR_d^{\mathrm{comm}}
+
SR_d^{\mathrm{primary}}
\nonumber\\
&\quad
+
\sum_{k\in\mathcal{S}_{\mathcal F}}
a_kx_{k,d}
\leq R_d,
\quad
\forall d\in\mathcal{D}_s
\label{eq:c_ram}\\
&
\sum_{c\in\mathcal{C}_d^{\mathcal F}}
z_{d,c}E_{d,c}^{\Delta,p}
+
\sum_{k\in\mathcal{S}}
x_{k,d}E_{k,d}^{\mathrm{comp}}
\nonumber\\
&\quad
+
\sum_{k=1}^{K}
\sum_{j\in\mathcal{D}_s}
\sum_{p\in\widetilde{\mathcal{P}}_{d,j}}
y_{k,d,j}^{p}
P_d^{\mathrm{tx},p}
T_{k,d,j}^{\mathrm{tx},p}
\nonumber\\
&\quad
+
\sum_{k=1}^{K-1}
\sum_{i\in\mathcal{D}_s}
\sum_{p\in\widetilde{\mathcal{P}}_{i,d}}
y_{k,i,d}^{p}
P_d^{\mathrm{rx},p}
T_{k,i,d}^{\mathrm{rx},p}
\leq E_d^{\mathrm{rem}},
\quad
\forall d\in\mathcal{D}_s
\label{eq:c_energy}\\
&
\sum_{j\in\mathcal{D}_s}
\sum_{p\in\widetilde{\mathcal{P}}_{i,j}}
y_{k,i,j}^{p}
=
x_{k,i},
\nonumber\\[-1mm]
&\hspace{35mm}
\forall k<K,\ \forall i\in\mathcal{D}_s
\label{eq:couple1}\\
&
\sum_{i\in\mathcal{D}_s}
\sum_{p\in\widetilde{\mathcal{P}}_{i,j}}
y_{k,i,j}^{p}
=
x_{k+1,j},
\nonumber\\[-1mm]
&\hspace{35mm}
\forall k<K,\ \forall j\in\mathcal{D}_s
\label{eq:couple2}\\
& x_{k,d},\;
y_{k,i,j}^{p},\;
z_{d,c}
\in\{0,1\}. \label{eq:binary}
\label{eq:binary}
\end{align}
\end{subequations}


where \(x_{k,d}\), \(y_{k,i,j}^{p}\), and \(z_{d,c}\) denote the
binary placement, routing, and replica reconfiguration-selection variables,
respectively. They are defined as
\begin{align}
x_{k,d}
 &= \mathbbm{1}\{\cdot\}\{\text{segment } k \text{ executes on device } d\},
\label{eq:x_definition}\\
y_{k,i,j}^{p}
&= \mathbbm{1}\{A_k \text{ is sent } i\!\to\!j \text{ via } p\}
\quad
p\in\widetilde{\mathcal{P}}_{i,j},
\label{eq:y_definition}\\
z_{d,c}
&=
\mathbbm{1}
\left\{
\text{replica reconfiguration }c
\text{ is selected for device }d
\right\},
\quad
c\in\mathcal{C}_d^{\mathcal{F}}.
\label{eq:z_definition}
\end{align}

}
{\color{orange}
Two consecutive model segments may be assigned either to different
devices or to the same device. If segments $k$ and $k+1$ execute on
different devices, the intermediate activation $\mathbf{A}_k$ is
transmitted using a wireless communication protocol. If they execute on the same
device, $\mathbf{A}_k$ is passed locally without wireless communication.

$\mathcal{P}_{i,j}$ denote the set of wireless protocols supported
between two distinct devices $i$ and $j$. To represent both wireless
and local transfers, we define the extended set of feasible transfer
modes as

\begin{equation}
\widetilde{\mathcal{P}}_{i,j}
=
\begin{cases}
\mathcal{P}_{i,j}, & i \neq j,\\
\{\mathrm{loc}\}, & i=j,
\end{cases}
\end{equation}

For a local transfer, no wireless transmission latency or radio energy
consumption is introduced. Therefore, we define

\begin{equation}
T_{k,d,d}^{\mathrm{loc}}=0,
\qquad
E_{k,d,d}^{\mathrm{loc}}=0,
\qquad
\forall k<K,\; d\in\mathcal{D}_s.
\end{equation}
}

{\color{orange}
The constraint in~\eqref{eq:fixed_primary} has been added to keep all
unaffected primary segments fixed at their original devices during
reconfiguration. Consequently, only the segments in $\mathcal{S}_F$,
whose primary devices have failed, can be reassigned to surviving devices.
}

\section{Solution}\label{sec:solution}

\newpage

\section{Results}\label{sec:results}

\ifflase

*-----------------SIMULATIONS----------------*

\subsection{Experimental Setup}

To evaluate the proposed \cgls{milp}-based replica-placement approach, a series of computational experiments was conducted and compared against the Random Feasible, Computation-Latency Greedy, Worst-Fit Remaining-Energy, and Flash First-Fit Decreasing replica placement algorithms. All algorithms were evaluated using the same device characteristics, failure combinations, resource constraints, computation costs, and communication parameters to ensure a fair comparison.

Three system sizes were considered, consisting of \(N\in\{10,15,20\}\) initially active devices. For each system size, the number of simultaneously failed devices was varied from \(F=1\) to \(F=5\). Every distinct combination of \(F\) failed devices among the \(N\) active devices was evaluated. The total number of failure combinations for a given pair \((N,F)\) is calculated as

\begin{equation}
N_{\mathrm{comb}}
=
\binom{N}{F}
=
\frac{N!}{F!(N-F)!}.
\label{eq:number_failure_combinations}
\end{equation}

The order of the failed devices does not affect the failure scenario. For example, the failure sets \(\{D_1,D_2\}\) and \(\{D_2,D_1\}\) represent the same failure combination and are therefore evaluated only once.

In the first scenario, \(N=10\) active devices were considered. This produced \(10\), \(45\), \(120\), \(210\), and \(252\) distinct failure combinations for \(F=1,2,3,4,\) and \(5\), respectively. Thus, \(637\) failure combinations were evaluated in total.

In the second scenario, the system contained \(N=15\) active devices. The corresponding numbers of failure combinations were \(15\), \(105\), \(455\), \(1{,}365\), and \(3{,}003\) for \(F=1,2,3,4,\) and \(5\), respectively. This resulted in a total of \(4{,}943\) failure combinations.

In the third scenario, \(N=20\) active devices were considered. The experiment evaluated \(20\), \(190\), \(1{,}140\), \(4{,}845\), and \(15{,}504\) failure combinations for \(F=1,2,3,4,\) and \(5\), respectively. This resulted in a total of \(21{,}699\) failure combinations.

\subsection{Evaluation Metrics}

The performance of the considered algorithms was assessed using the following metrics:

\begin{itemize}
    \item \textbf{End-to-end inference latency:}
    This metric measures the total post-reconfiguration latency, including segment computation latency and intermediate-activation transfer latency. The latency distributions are reported over the failure combinations for which the corresponding algorithm found a feasible solution.

    \item \textbf{Resource-utilization distribution:}
    This metric shows the flash, SRAM, and energy utilization of the surviving devices across all feasible failure combinations. It illustrates how each replica-placement algorithm distributes the additional replica workload among the available devices.

    \item \textbf{Computation time:}
    This metric measures the time required by each algorithm to construct and evaluate a solution for one failure combination. For each value of \(F\), the median computation time is reported, while the shaded region represents the interquartile range across all evaluated failure combinations.

    \item \textbf{Feasibility rate:}
    This metric represents the percentage of failure combinations for which an algorithm found a placement satisfying all flash, SRAM, energy, routing, and communication-protocol constraints. It is calculated as

    \begin{equation}
    \mathrm{Feasibility\ Rate}
    =
    \frac{N_{\mathrm{feasible}}}
         {N_{\mathrm{total}}}
    \times 100\%.
    \label{eq:feasibility_rate}
    \end{equation}
\end{itemize}

These figures jointly evaluate the solution quality, resource utilization, computational cost, and robustness of the algorithms as the number of simultaneous device failures increases.

\begin{figure*}[!t]
    \centering

    \subfloat[\(N=10\) active devices]{
        \includegraphics[width=0.47\textwidth]
        {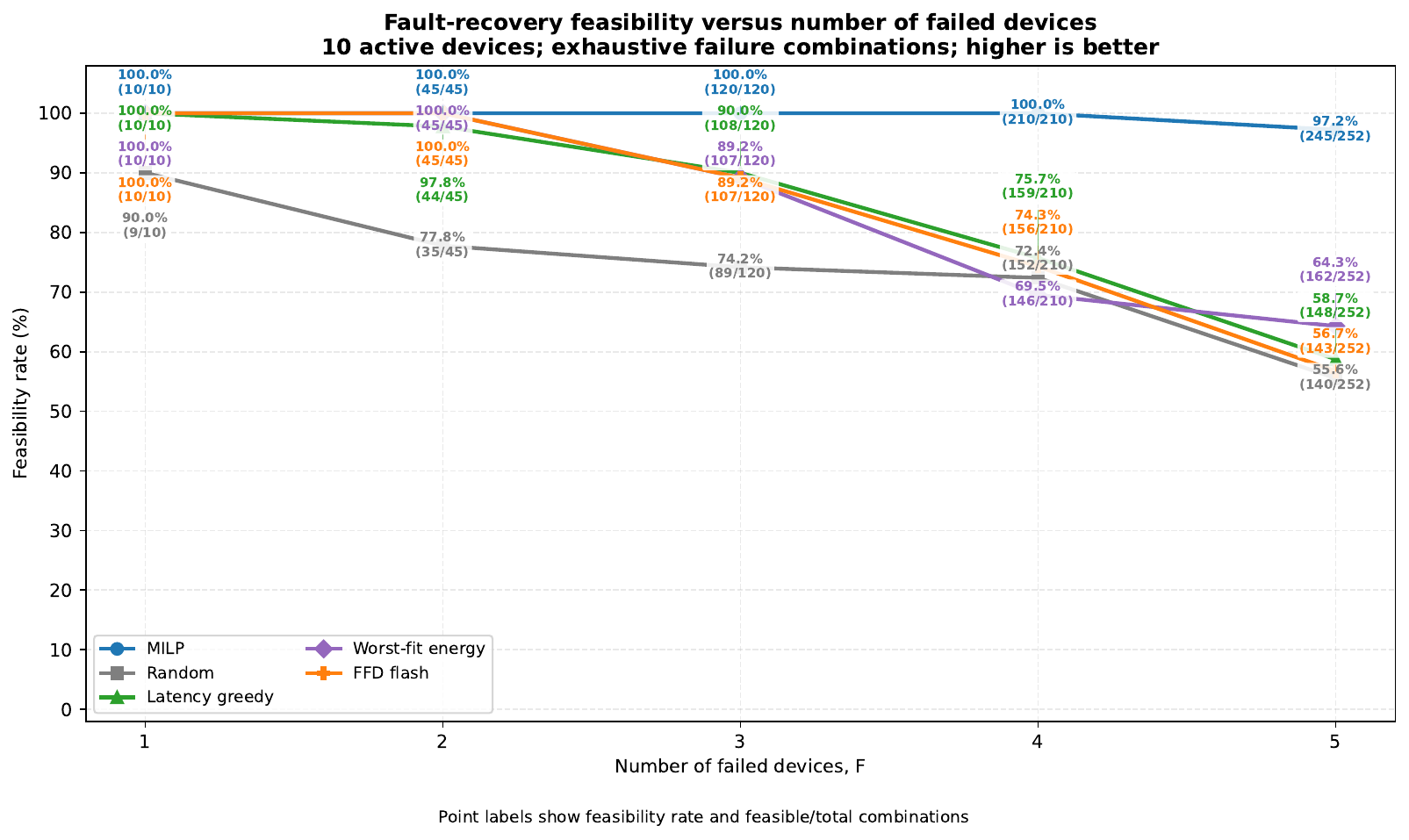}
        \label{fig:feasibility_n10}
    }
    \hfill
    \subfloat[\(N=15\) active devices]{
        \includegraphics[width=0.47\textwidth]
        {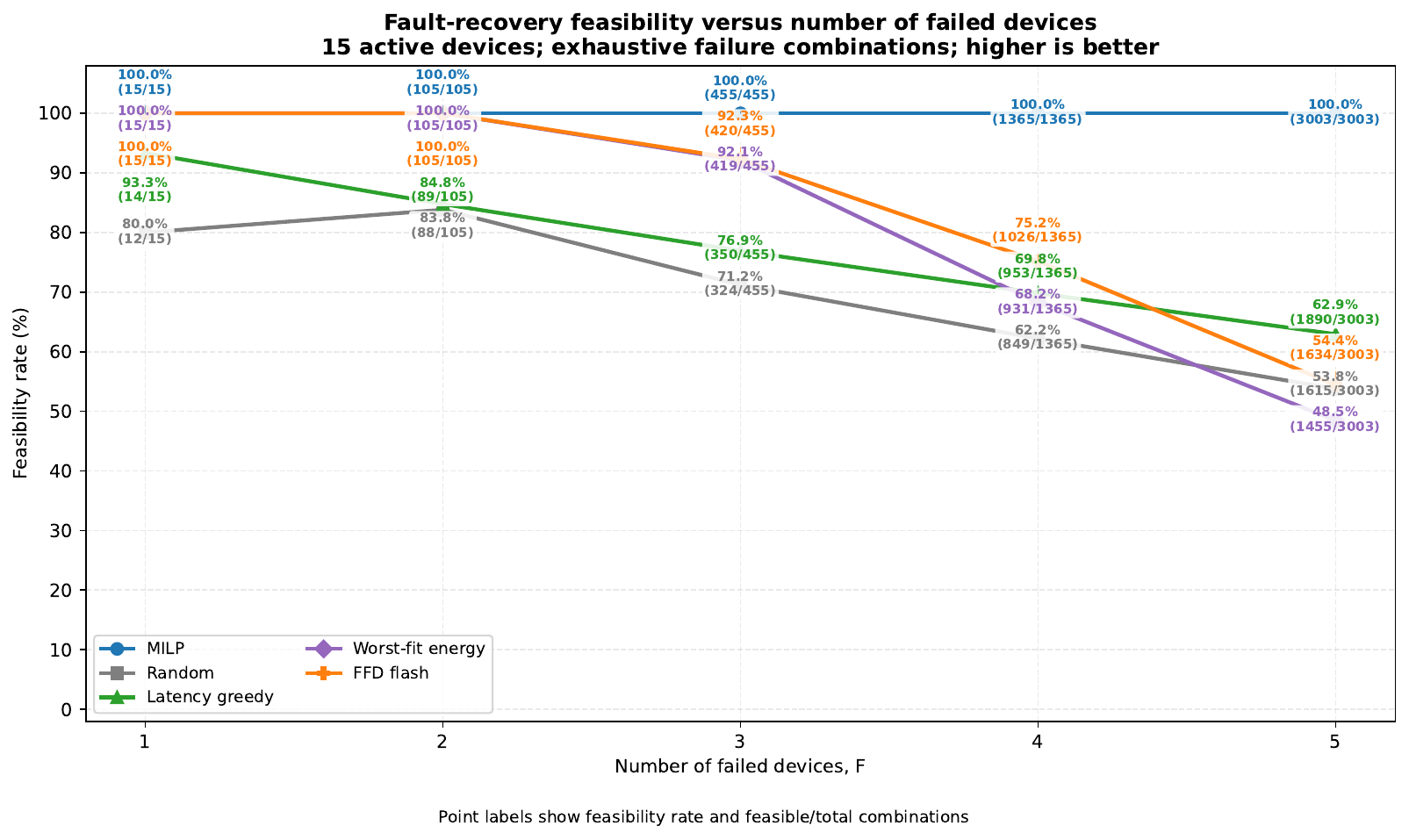}
        \label{fig:feasibility_n15}
    }

    \vspace{0.3cm}

    \subfloat[\(N=20\) active devices]{
        \includegraphics[width=0.47\textwidth]
        {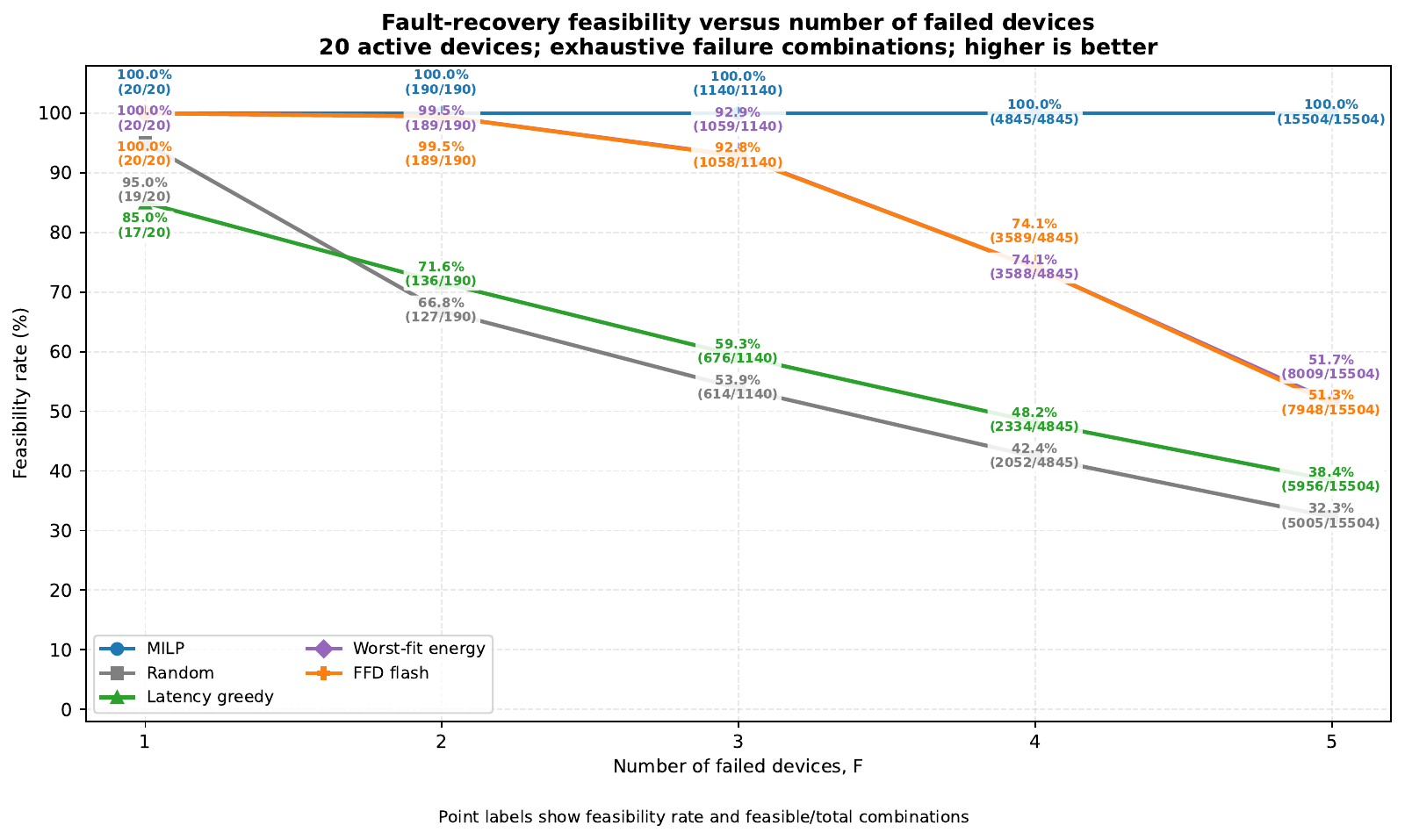}
        \label{fig:feasibility_n20}
    }

    \caption{Feasibility rate versus the number of failed devices for
    (a) \(N=10\), (b) \(N=15\), and (c) \(N=20\) initially active
    devices. Each point represents the percentage of feasible solutions
    over all \(\binom{N}{F}\) failure combinations.}

    \label{fig:feasibility_all_devices}
\end{figure*}

\begin{figure*}[!t]
    \centering

    \subfloat[\(N=10\) active devices
    \label{fig:computation_time_n10}]{
        \includegraphics[width=0.47\textwidth]
        {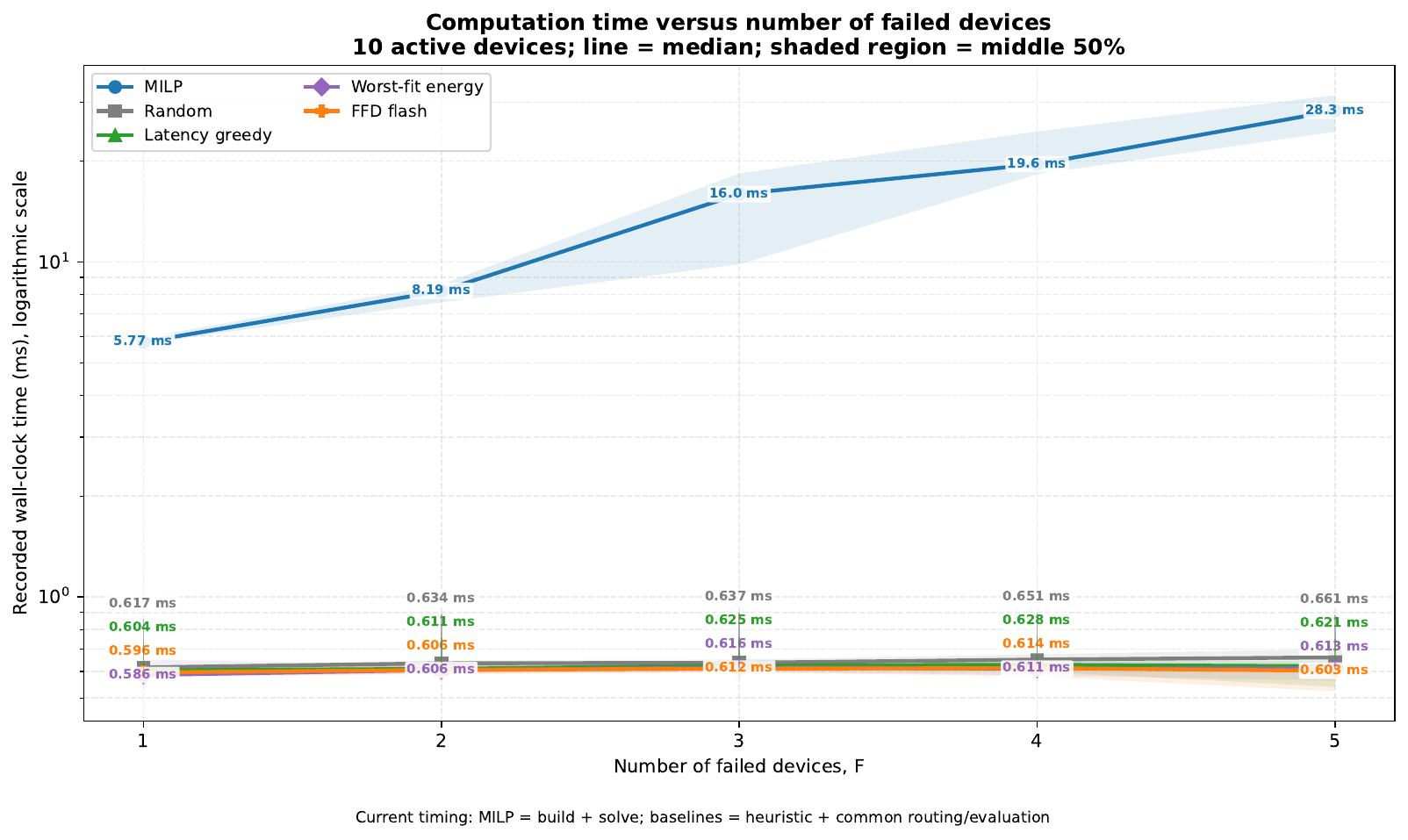}
    }
    \hfill
    \subfloat[\(N=15\) active devices
    \label{fig:computation_time_n15}]{
        \includegraphics[width=0.47\textwidth]
        {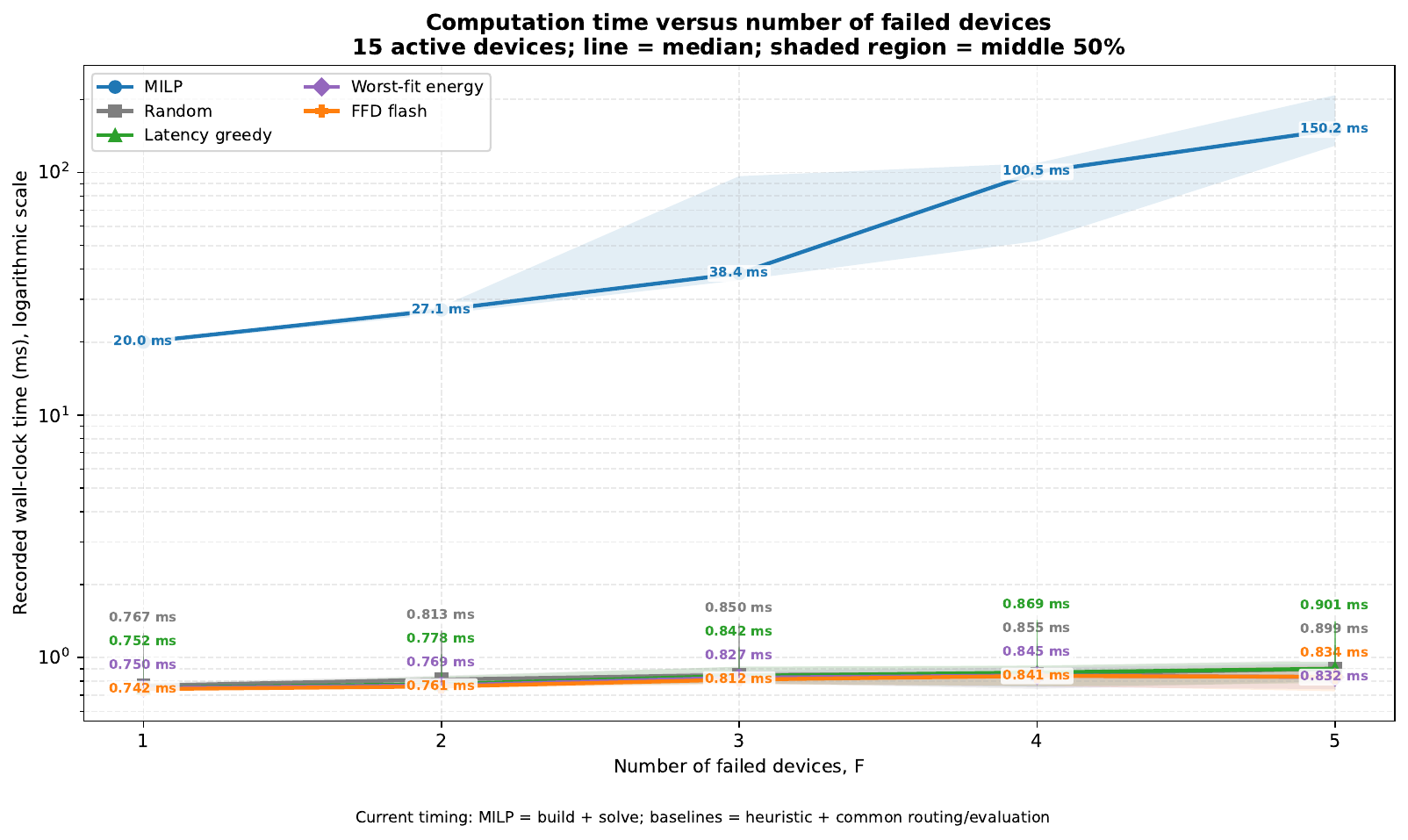}
    }

    \par\medskip

    \subfloat[\(N=20\) active devices
    \label{fig:computation_time_n20}]{
        \includegraphics[width=0.47\textwidth]
        {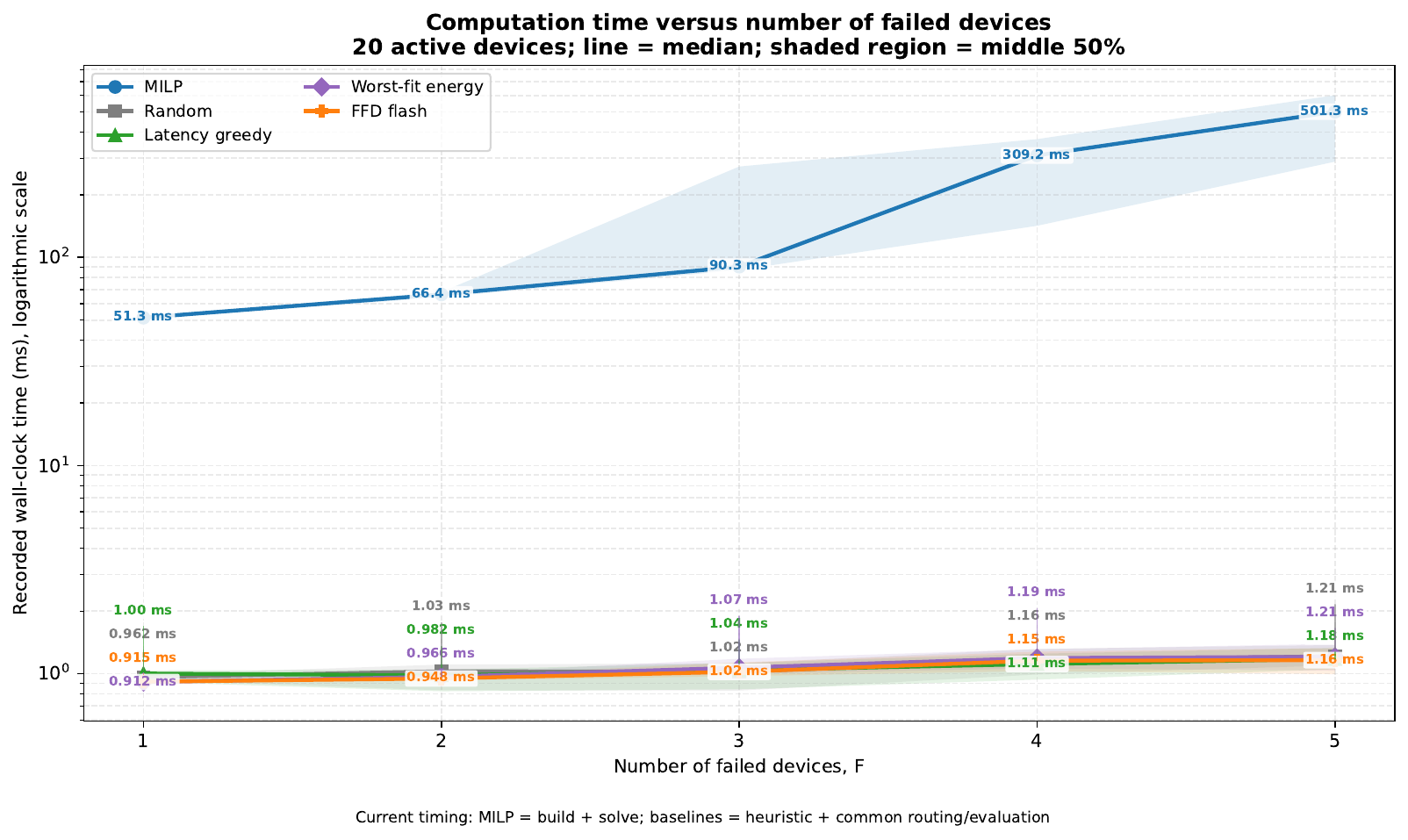}
    }

    \caption{computation time versus the number of failed
    devices for (a) \(N=10\), (b) \(N=15\), and (c) \(N=20\)
    initially active devices. For each value of \(F\), the line indicates
    the median computation time per failure combination, while the shaded
    region represents the interquartile range over all
    \(\binom{N}{F}\) evaluated failure combinations. The vertical axis
    uses a logarithmic scale, and lower values indicate faster execution.
    The \cgls{milp} timing includes model construction and optimization, whereas
    the baseline timing includes heuristic placement and the common
    routing and feasibility evaluation.}

    \label{fig:computation_time_all_devices}
\end{figure*}

\begin{figure*}[!t]
    \centering

    \subfloat[\(F=1\): \(10\) combinations
    \label{fig:e2e_n10_f1}]{
        \includegraphics[width=0.47\textwidth]
        {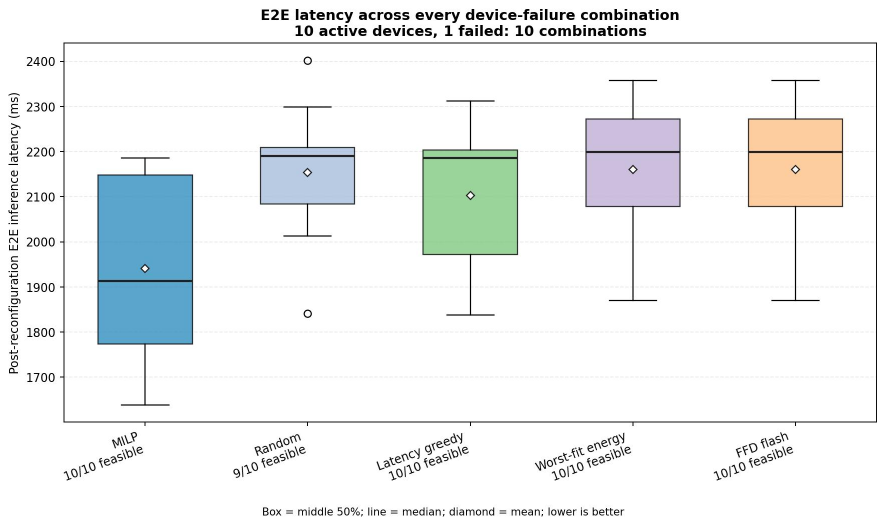}
    }
    \hfill
    \subfloat[\(F=2\): \(45\) combinations
    \label{fig:e2e_n10_f2}]{
        \includegraphics[width=0.47\textwidth]
        {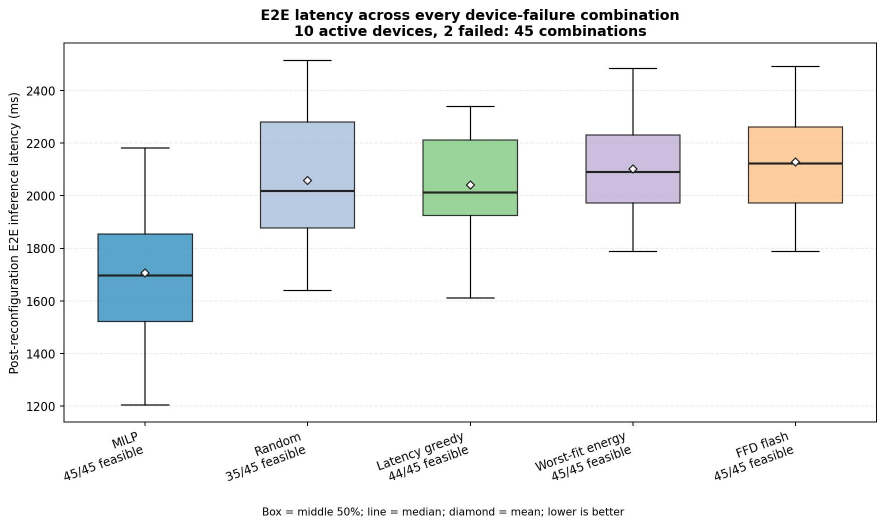}
    }

    \par\medskip

    \subfloat[\(F=3\): \(120\) combinations
    \label{fig:e2e_n10_f3}]{
        \includegraphics[width=0.47\textwidth]
        {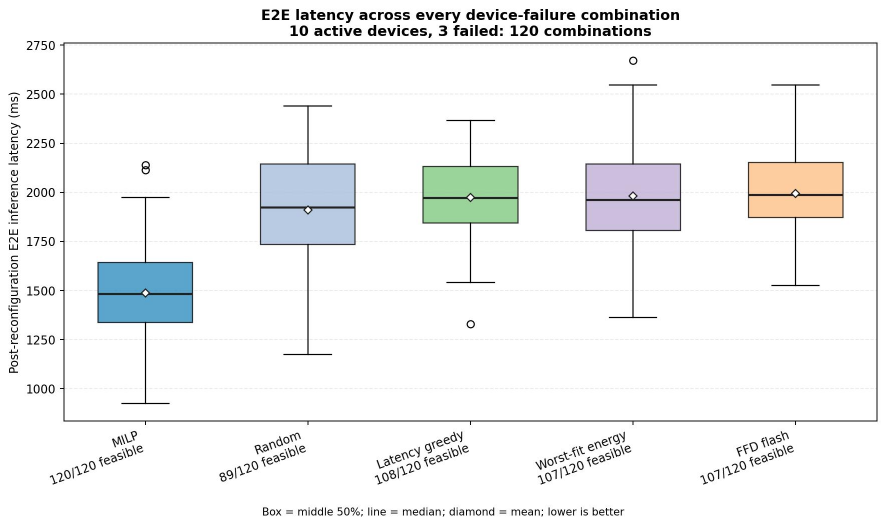}
    }
    \hfill
    \subfloat[\(F=4\): \(210\) combinations
    \label{fig:e2e_n10_f4}]{
        \includegraphics[width=0.47\textwidth]
        {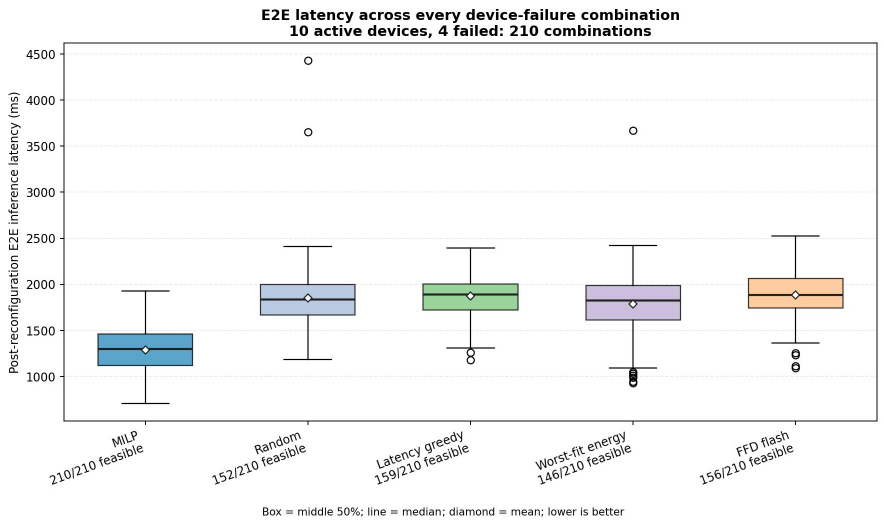}
    }

    \par\medskip

    \subfloat[\(F=5\): \(252\) combinations
    \label{fig:e2e_n10_f5}]{
        \includegraphics[width=0.47\textwidth]
        {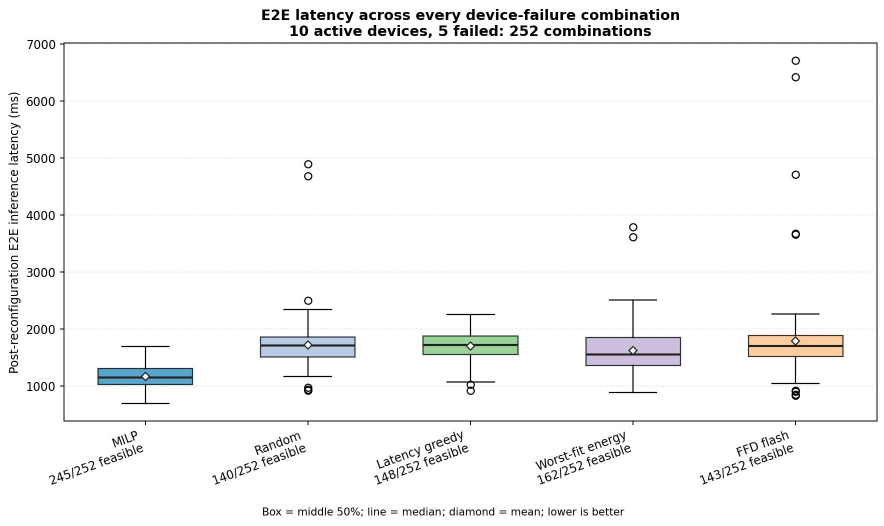}
    }

    \caption{Distribution of the post-reconfiguration E2E inference
    latency for \(N=10\) initially active devices under
    (a) \(F=1\), (b) \(F=2\), (c) \(F=3\), (d) \(F=4\), and
    (e) \(F=5\) simultaneous device failures. For each failure count,
    all \(\binom{10}{F}\) distinct failure combinations were evaluated.
    Each boxplot contains the E2E latencies of the failure combinations
    for which the corresponding algorithm returned a feasible solution.
    The center line represents the median, the diamond represents the
    mean, and lower latency values indicate better performance.}

    \label{fig:e2e_latency_n10}
\end{figure*}

\begin{figure*}[!t]
    \centering

    \subfloat[\(F=1\): \(15\) combinations
    \label{fig:e2e_n10_f1}]{
        \includegraphics[width=0.47\textwidth]
        {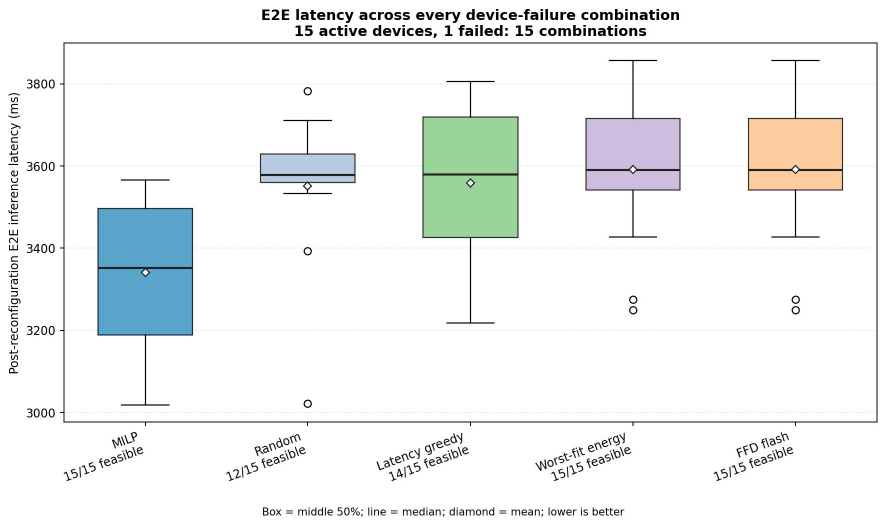}
    }
    \hfill
    \subfloat[\(F=2\): \(105\) combinations
    \label{fig:e2e_n10_f2}]{
        \includegraphics[width=0.47\textwidth]
        {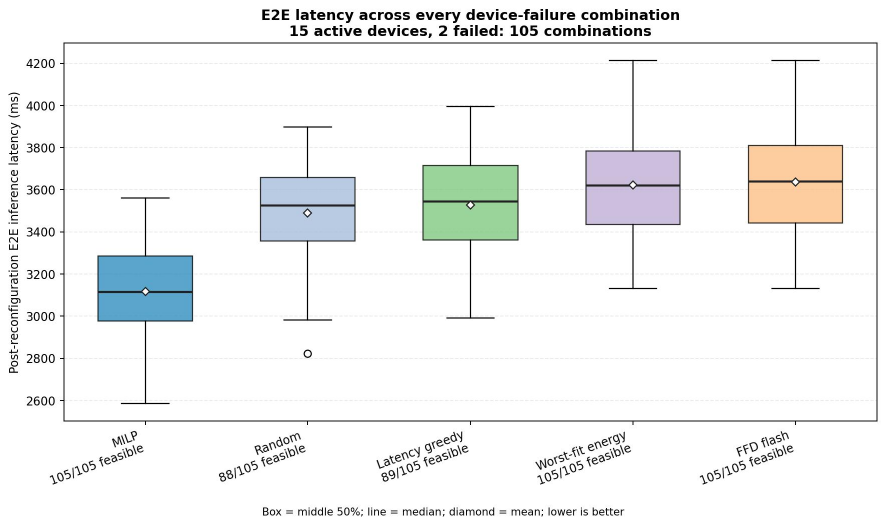}
    }

    \par\medskip

    \subfloat[\(F=3\): \(455\) combinations
    \label{fig:e2e_n10_f3}]{
        \includegraphics[width=0.47\textwidth]
        {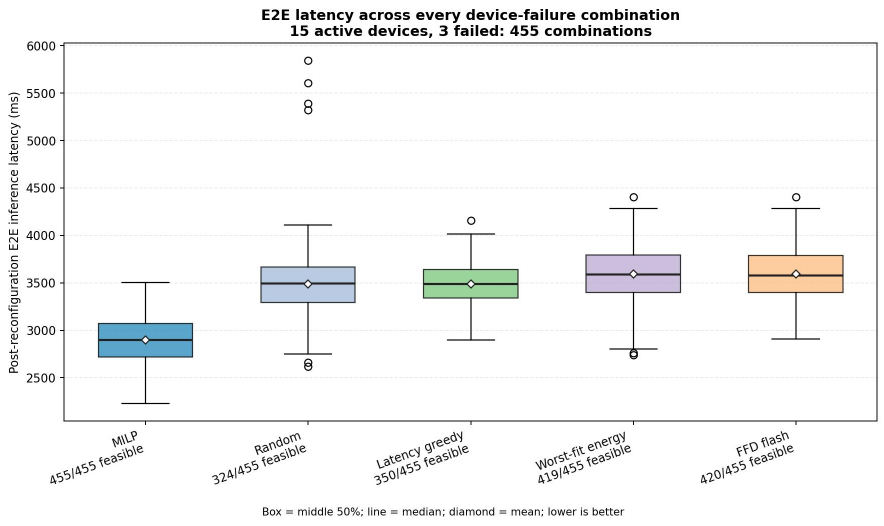}
    }
    \hfill
    \subfloat[\(F=4\): \(1365\) combinations
    \label{fig:e2e_n10_f4}]{
        \includegraphics[width=0.47\textwidth]
        {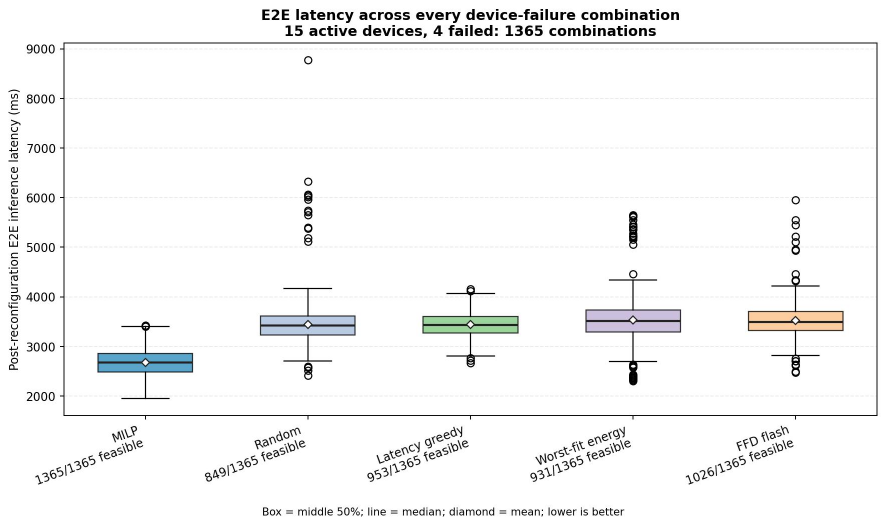}
    }

    \par\medskip

    \subfloat[\(F=5\): \(3003\) combinations
    \label{fig:e2e_n10_f5}]{
        \includegraphics[width=0.47\textwidth]
        {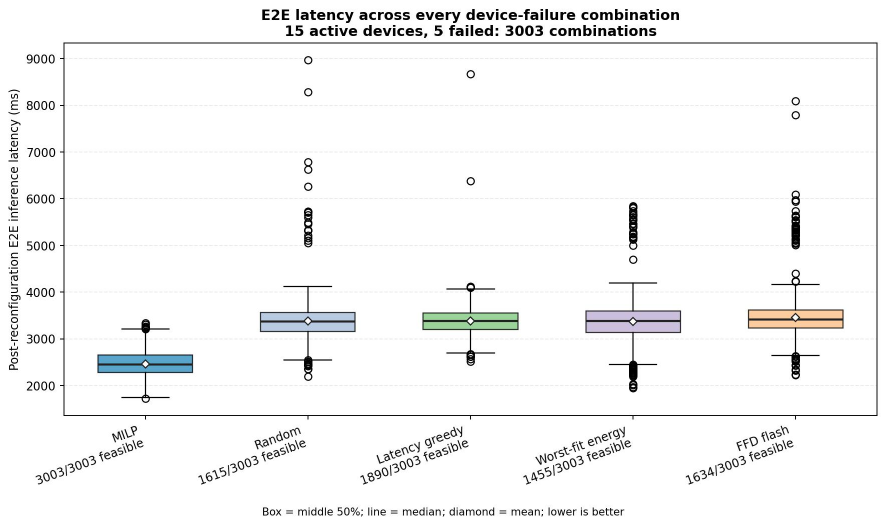}
    }

    \caption{Distribution of the post-reconfiguration E2E inference
    latency for \(N=15\) initially active devices under
    (a) \(F=1\), (b) \(F=2\), (c) \(F=3\), (d) \(F=4\), and
    (e) \(F=5\) simultaneous device failures. For each failure count,
    all \(\binom{10}{F}\) distinct failure combinations were evaluated.
    Each boxplot contains the E2E latencies of the failure combinations
    for which the corresponding algorithm returned a feasible solution.
    The center line represents the median, the diamond represents the
    mean, and lower latency values indicate better performance.}

    \label{fig:e2e_latency_n10}
\end{figure*}

\begin{figure*}[!t]
    \centering

    \subfloat[\(F=1\): \(20\) combinations
    \label{fig:e2e_n10_f1}]{
        \includegraphics[width=0.47\textwidth]
        {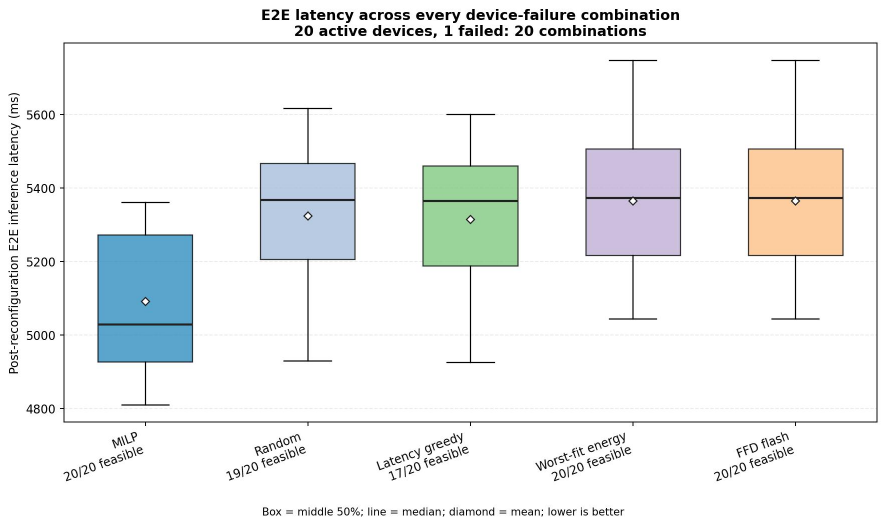}
    }
    \hfill
    \subfloat[\(F=2\): \(190\) combinations
    \label{fig:e2e_n10_f2}]{
        \includegraphics[width=0.47\textwidth]
        {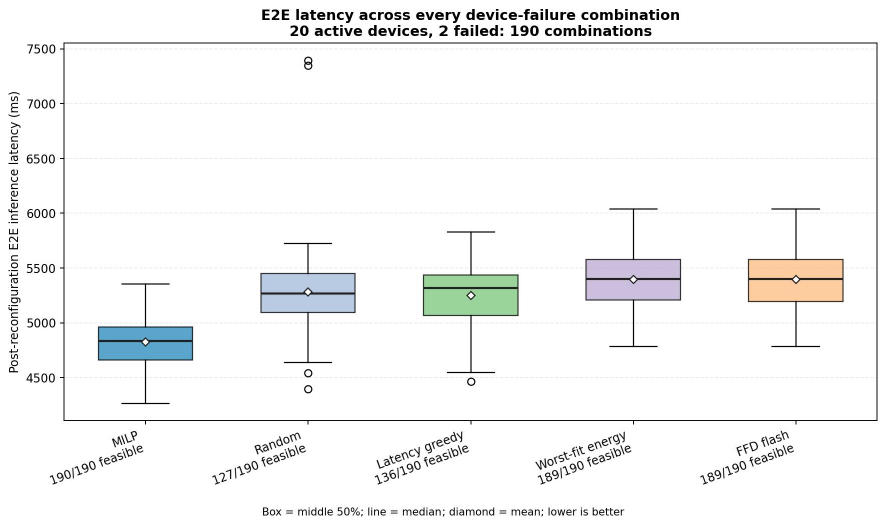}
    }

    \par\medskip

    \subfloat[\(F=3\): \(1140\) combinations
    \label{fig:e2e_n10_f3}]{
        \includegraphics[width=0.47\textwidth]
        {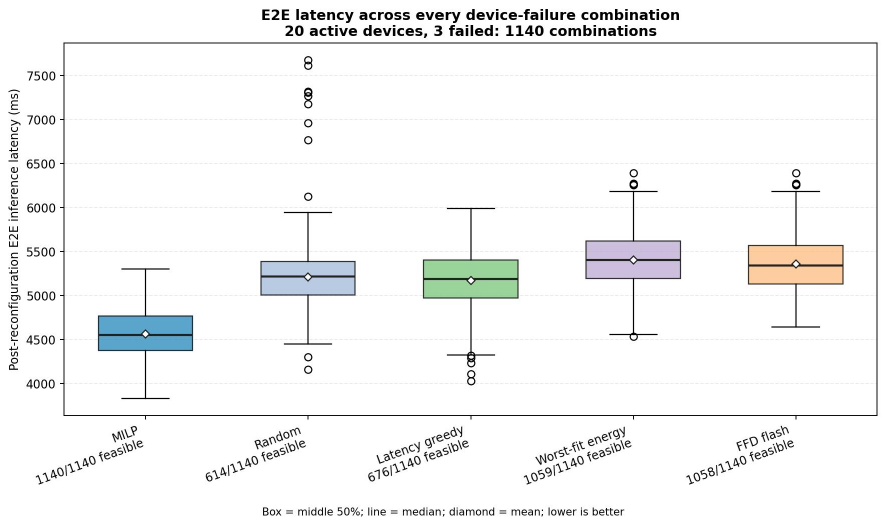}
    }
    \hfill
    \subfloat[\(F=4\): \(4845\) combinations
    \label{fig:e2e_n10_f4}]{
        \includegraphics[width=0.47\textwidth]
        {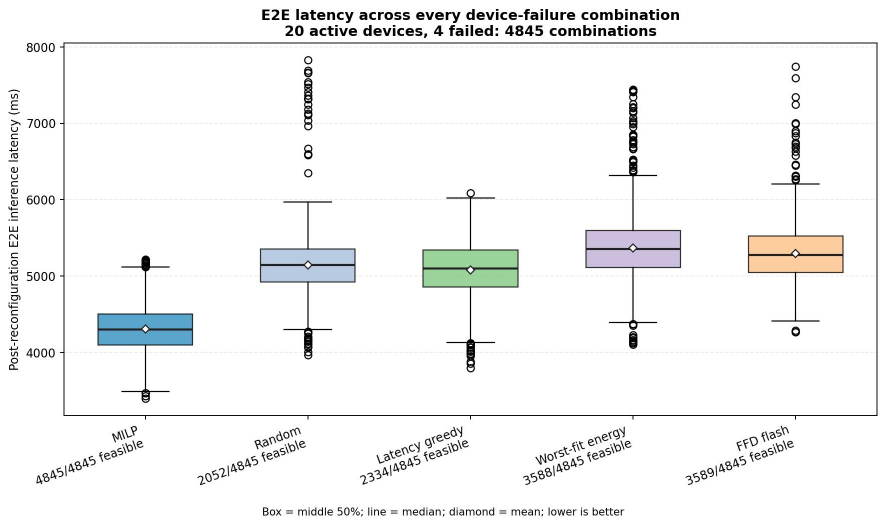}
    }

    \par\medskip

    \subfloat[\(F=5\): \(15504\) combinations
    \label{fig:e2e_n10_f5}]{
        \includegraphics[width=0.47\textwidth]
        {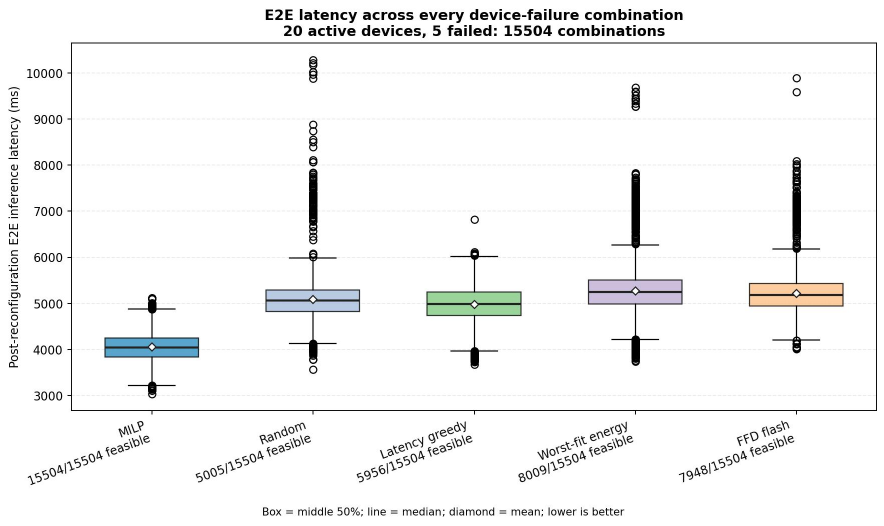}
    }

    \caption{Distribution of the post-reconfiguration E2E inference
    latency for \(N=20\) initially active devices under
    (a) \(F=1\), (b) \(F=2\), (c) \(F=3\), (d) \(F=4\), and
    (e) \(F=5\) simultaneous device failures. For each failure count,
    all \(\binom{10}{F}\) distinct failure combinations were evaluated.
    Each boxplot contains the E2E latencies of the failure combinations
    for which the corresponding algorithm returned a feasible solution.
    The center line represents the median, the diamond represents the
    mean, and lower latency values indicate better performance.}

    \label{fig:e2e_latency_n10}
\end{figure*}

\begin{figure*}[!t]
    \centering

    \subfloat[\(F=1\): \(10\) combinations
    \label{fig:resource_n20_f1}]{
        \includegraphics[width=0.47\textwidth]
        {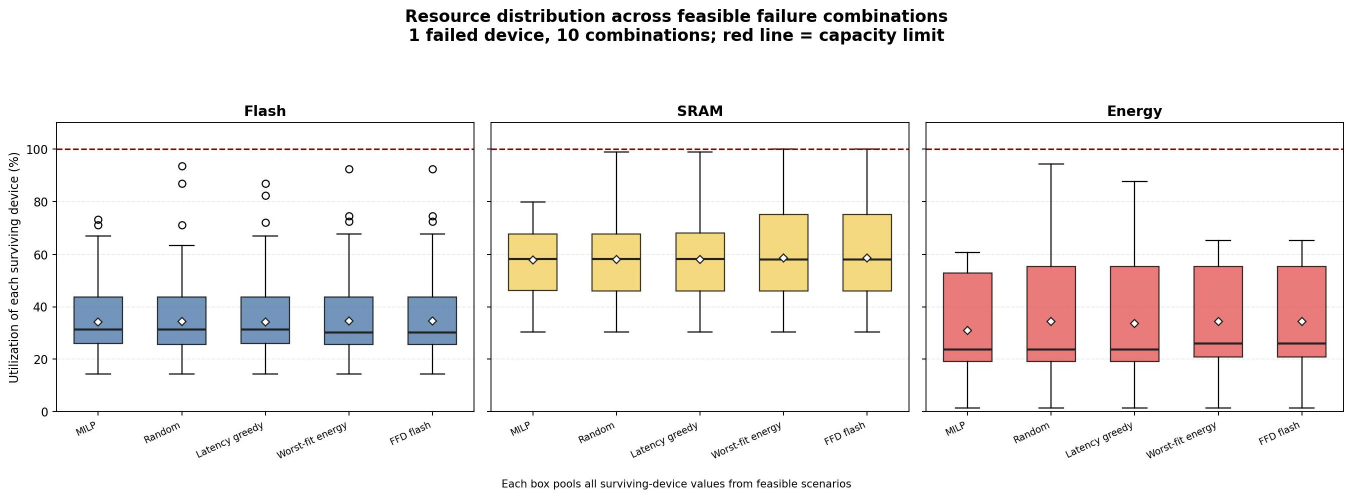}
    }
    \hfill
    \subfloat[\(F=2\): \(45\) combinations
    \label{fig:resource_n20_f2}]{
        \includegraphics[width=0.47\textwidth]
        {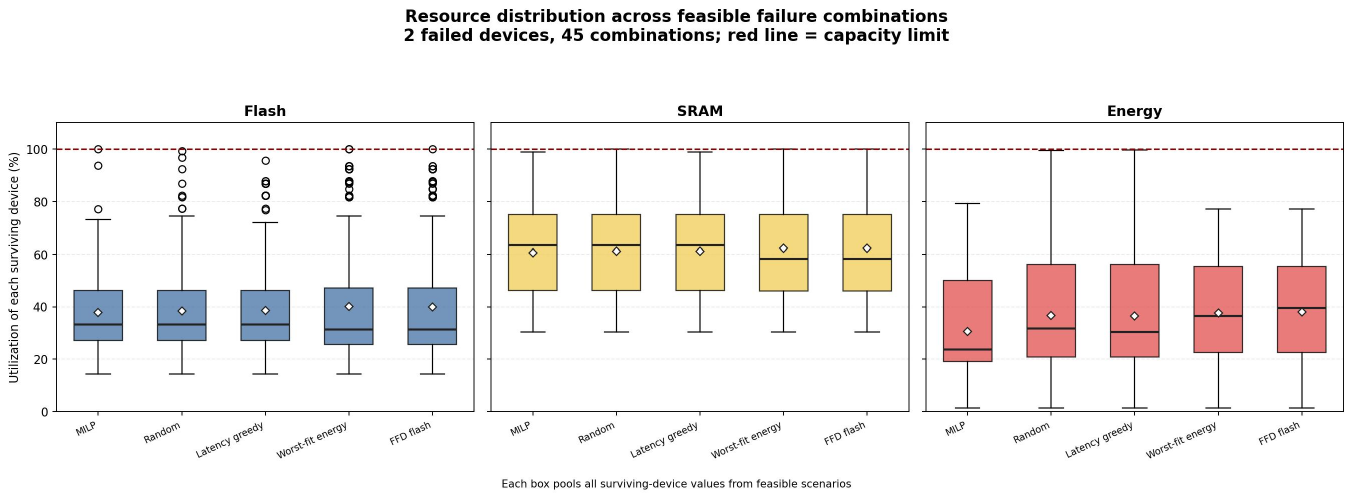}
    }

    \par\medskip

    \subfloat[\(F=3\): \(120\) combinations
    \label{fig:resource_n20_f3}]{
        \includegraphics[width=0.47\textwidth]
        {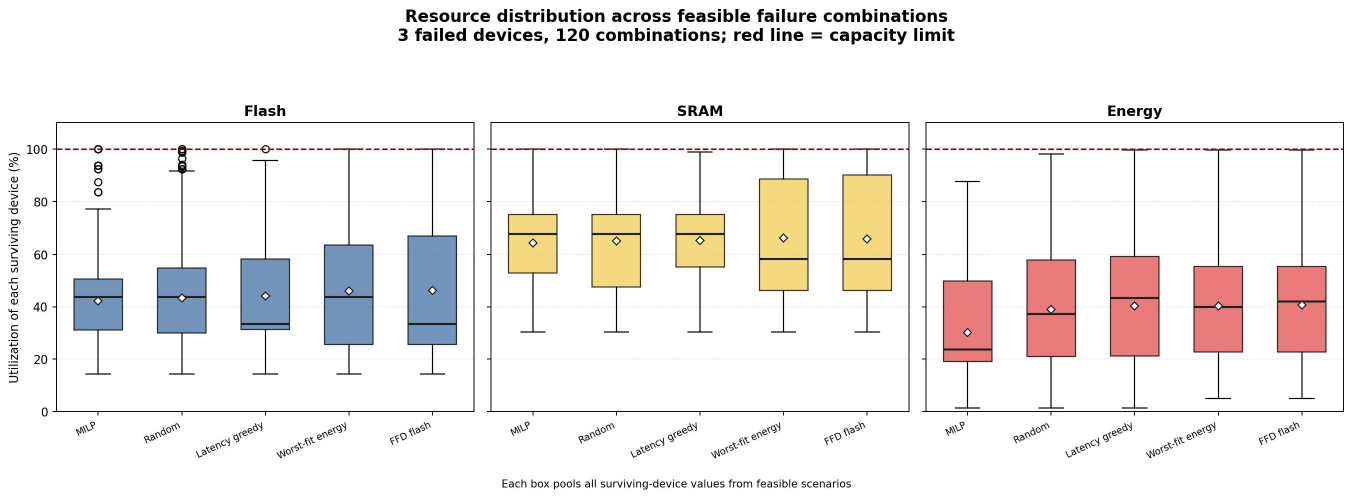}
    }
    \hfill
    \subfloat[\(F=4\): \(210\) combinations
    \label{fig:resource_n20_f4}]{
        \includegraphics[width=0.47\textwidth]
        {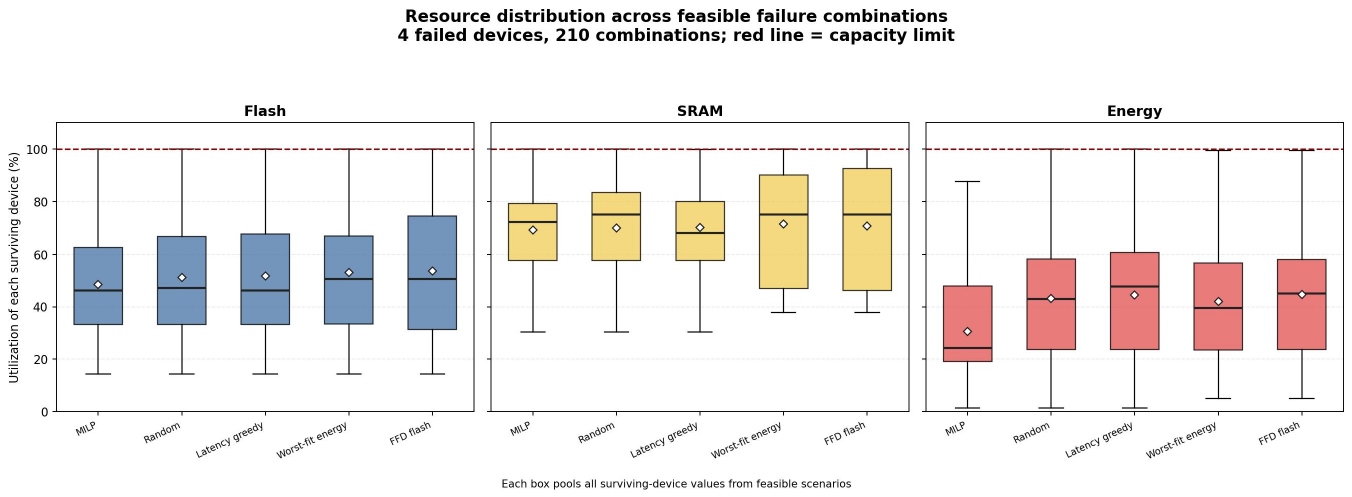}
    }

    \par\medskip

    \subfloat[\(F=5\): \(252\) combinations
    \label{fig:resource_n20_f5}]{
        \includegraphics[width=0.47\textwidth]
        {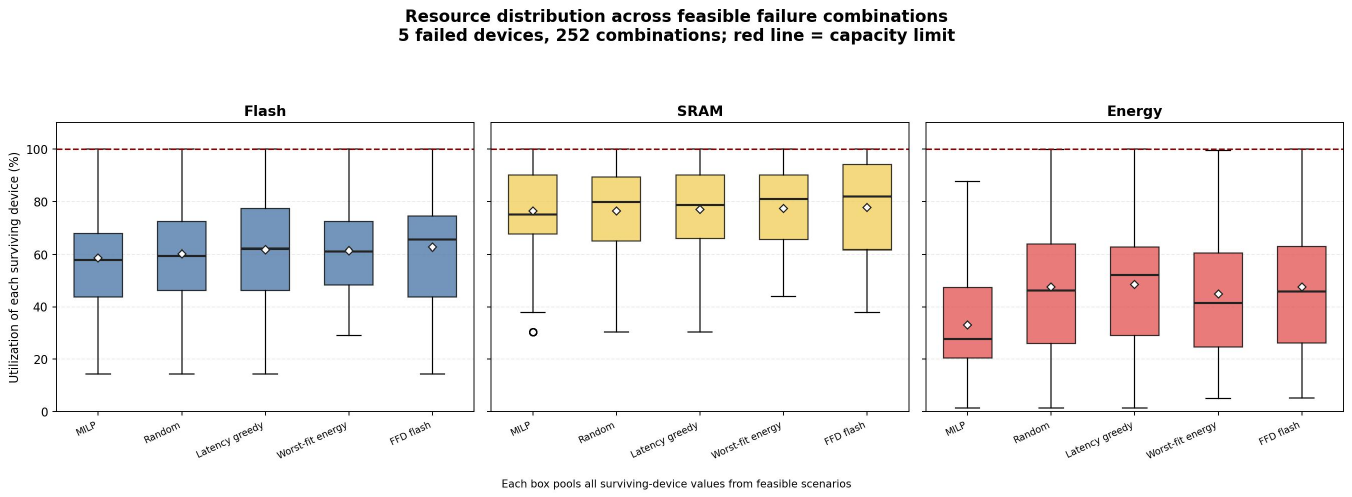}
    }

    \caption{Distribution of flash, SRAM, and energy utilization across
    the surviving devices for \(N=10\) initially active devices under
    (a) \(F=1\), (b) \(F=2\), (c) \(F=3\), (d) \(F=4\), and
    (e) \(F=5\) simultaneous device failures. For each failure count,
    all \(\binom{20}{F}\) distinct failure combinations were evaluated.
    Each boxplot pools the device-level utilization observations from
    the failure combinations for which the corresponding algorithm
    returned a feasible solution. The center line represents the median,
    the diamond represents the mean, the whiskers extend to
    \(1.5\times IQR\), and circles represent outliers. The red dashed
    line denotes the resource-capacity limit of \(100\%\). Lower values
    indicate greater remaining resource headroom.}

    \label{fig:resource_utilization_n20}
\end{figure*}

\begin{figure*}[!t]
    \centering

    \subfloat[\(F=1\): \(15\) combinations
    \label{fig:resource_n20_f1}]{
        \includegraphics[width=0.47\textwidth]
        {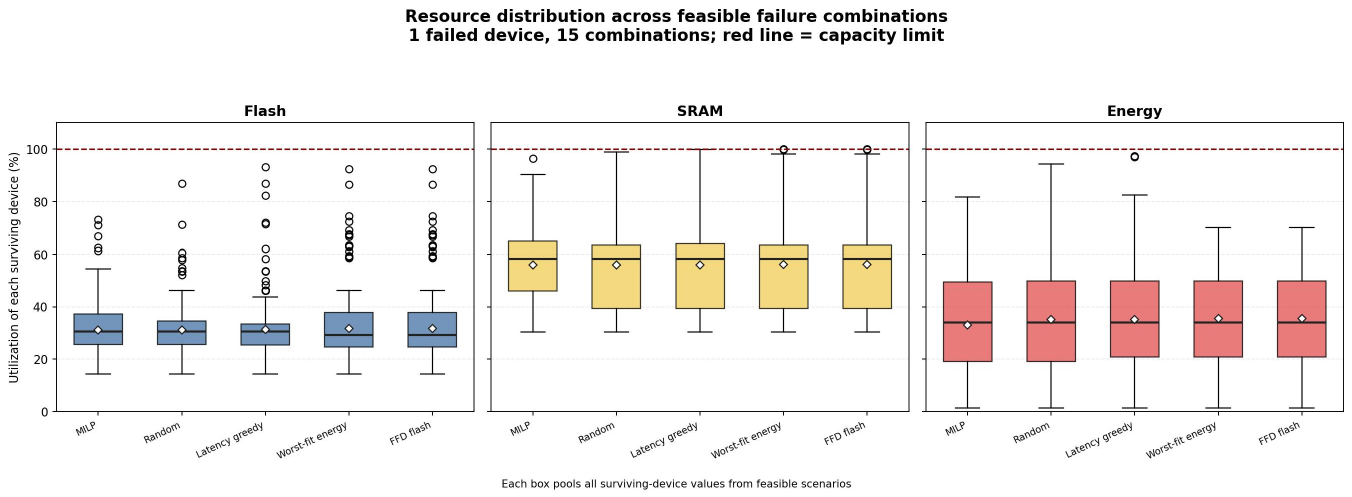}
    }
    \hfill
    \subfloat[\(F=2\): \(105\) combinations
    \label{fig:resource_n20_f2}]{
        \includegraphics[width=0.47\textwidth]
        {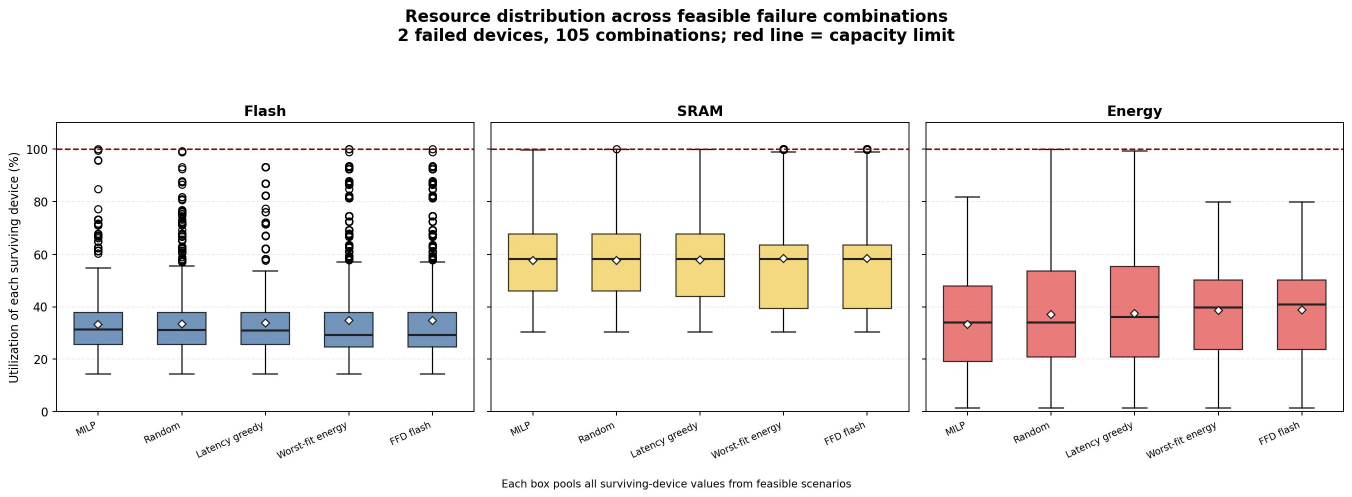}
    }

    \par\medskip

    \subfloat[\(F=3\): \(455\) combinations
    \label{fig:resource_n20_f3}]{
        \includegraphics[width=0.47\textwidth]
        {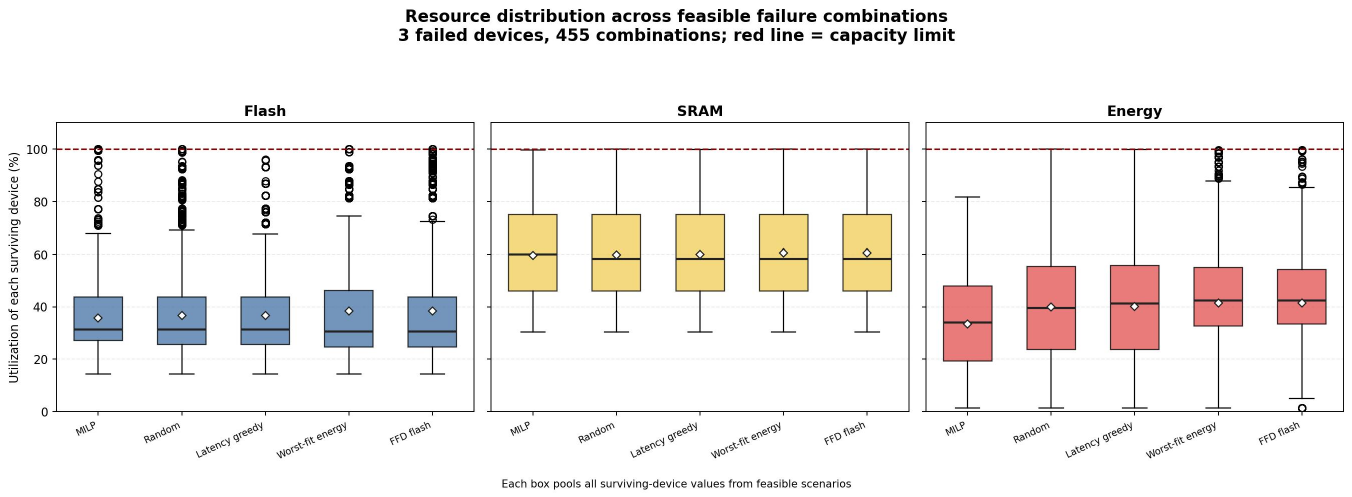}
    }
    \hfill
    \subfloat[\(F=4\): \(1365\) combinations
    \label{fig:resource_n20_f4}]{
        \includegraphics[width=0.47\textwidth]
        {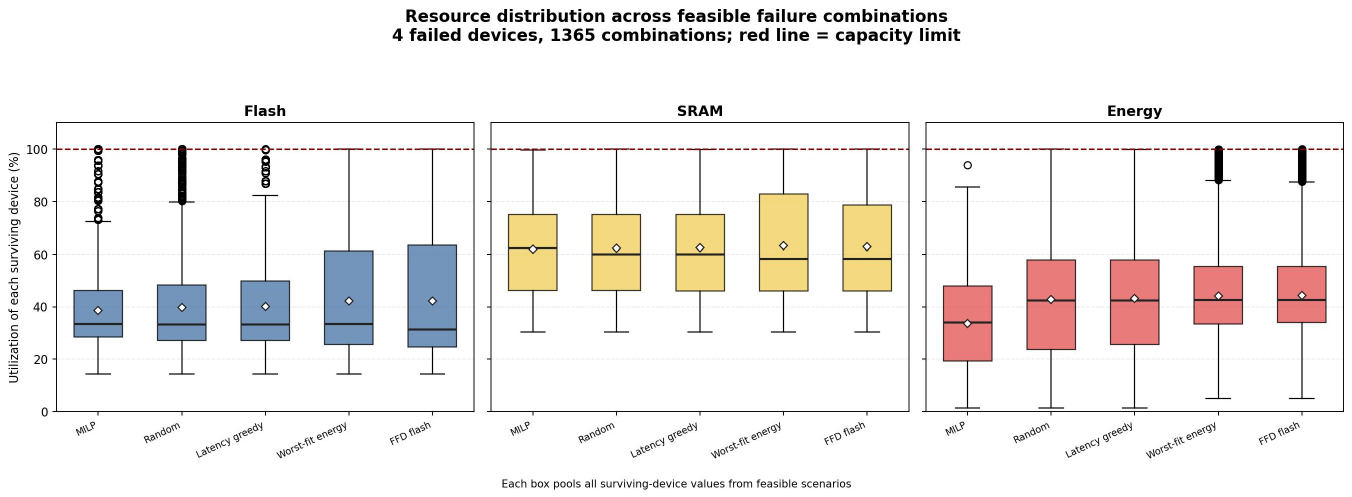}
    }

    \par\medskip

    \subfloat[\(F=5\): \(3003\) combinations
    \label{fig:resource_n20_f5}]{
        \includegraphics[width=0.47\textwidth]
        {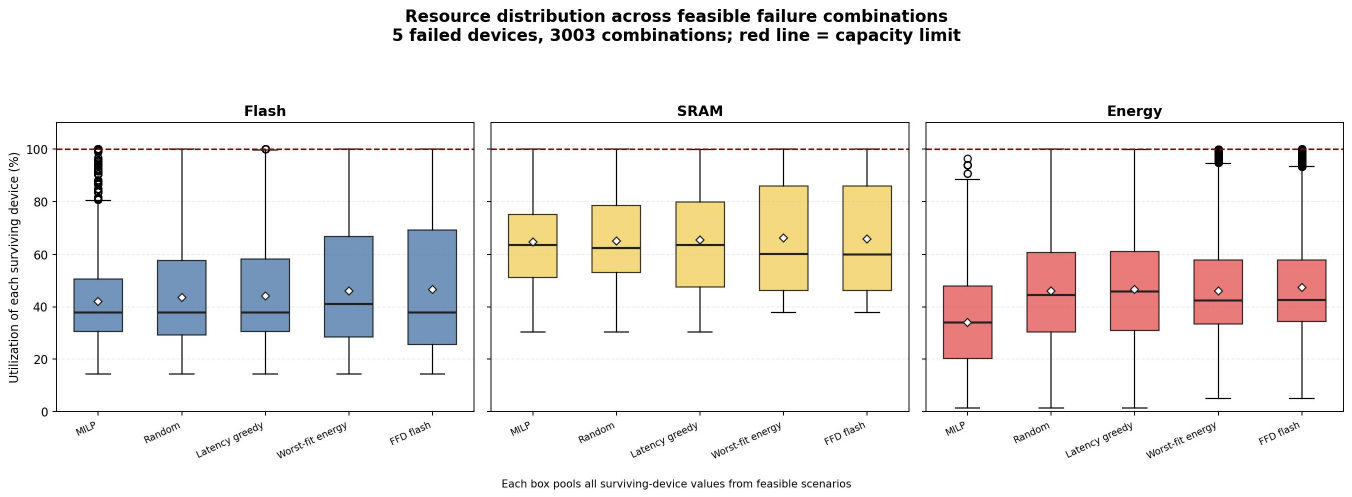}
    }

    \caption{Distribution of flash, SRAM, and energy utilization across
    the surviving devices for \(N=15\) initially active devices under
    (a) \(F=1\), (b) \(F=2\), (c) \(F=3\), (d) \(F=4\), and
    (e) \(F=5\) simultaneous device failures. For each failure count,
    all \(\binom{20}{F}\) distinct failure combinations were evaluated.
    Each boxplot pools the device-level utilization observations from
    the failure combinations for which the corresponding algorithm
    returned a feasible solution. The center line represents the median,
    the diamond represents the mean, the whiskers extend to
    \(1.5\times IQR\), and circles represent outliers. The red dashed
    line denotes the resource-capacity limit of \(100\%\). Lower values
    indicate greater remaining resource headroom.}

    \label{fig:resource_utilization_n20}
\end{figure*}

\begin{figure*}[!t]
    \centering

    \subfloat[\(F=1\): \(20\) combinations
    \label{fig:resource_n20_f1}]{
        \includegraphics[width=0.47\textwidth]
        {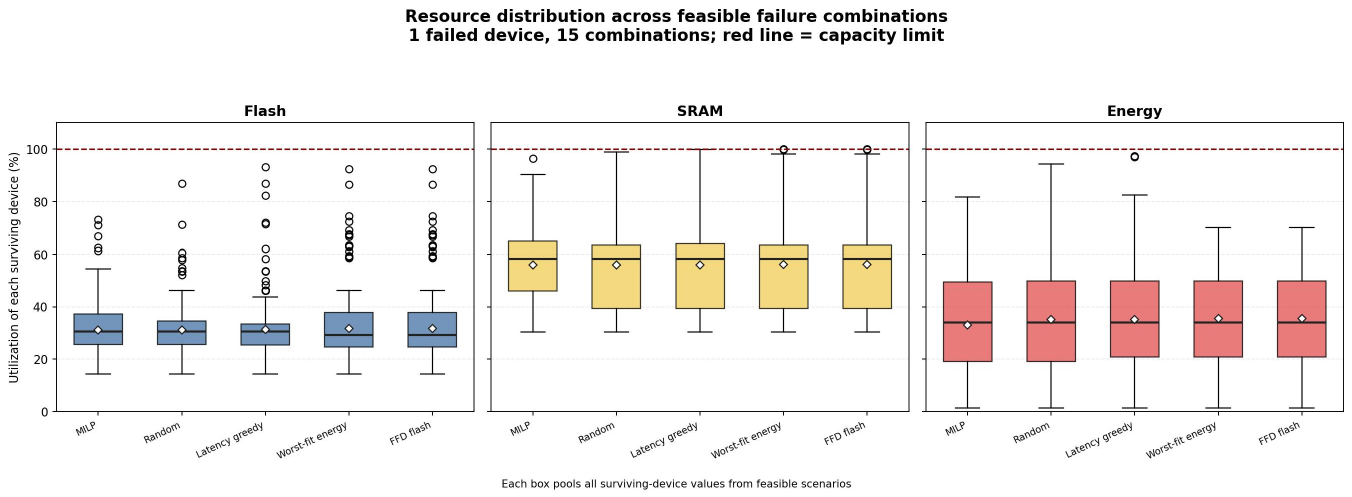}
    }
    \hfill
    \subfloat[\(F=2\): \(190\) combinations
    \label{fig:resource_n20_f2}]{
        \includegraphics[width=0.47\textwidth]
        {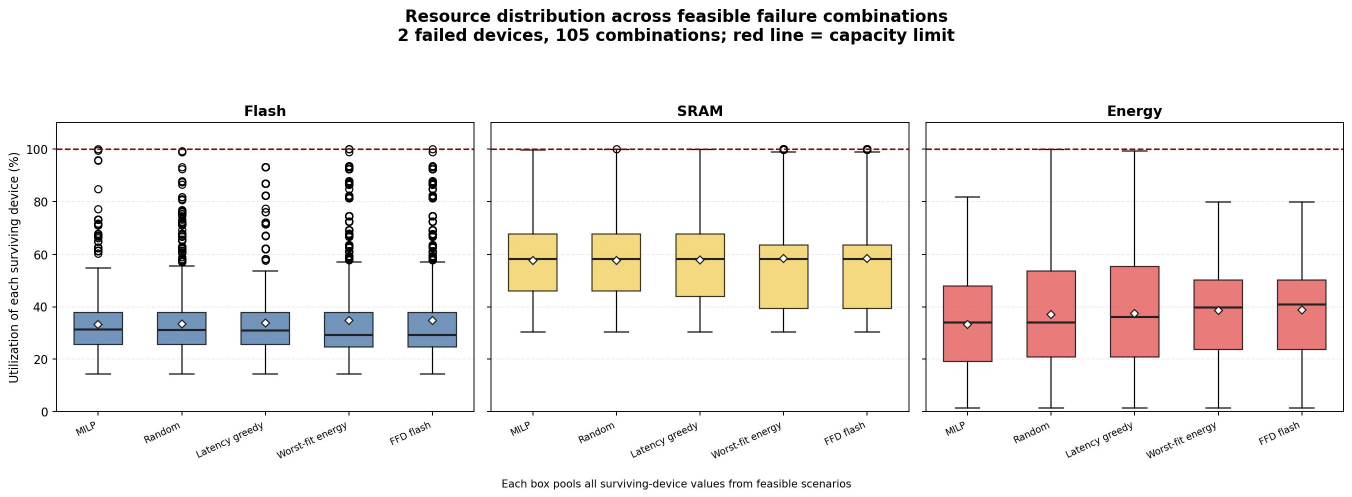}
    }

    \par\medskip

    \subfloat[\(F=3\): \(1140\) combinations
    \label{fig:resource_n20_f3}]{
        \includegraphics[width=0.47\textwidth]
        {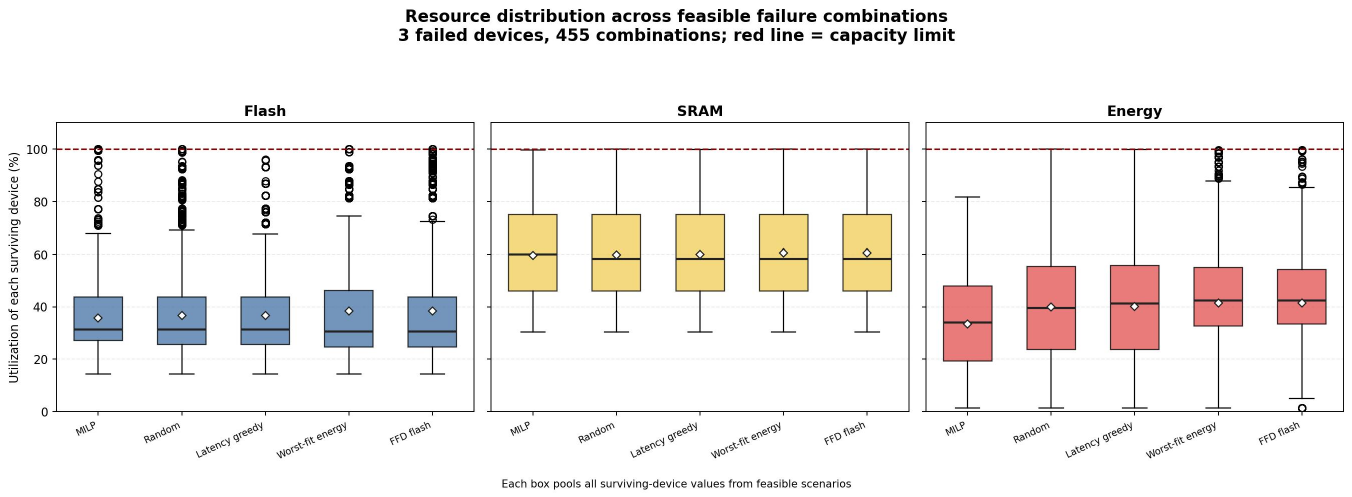}
    }
    \hfill
    \subfloat[\(F=4\): \(4845\) combinations
    \label{fig:resource_n20_f4}]{
        \includegraphics[width=0.47\textwidth]
        {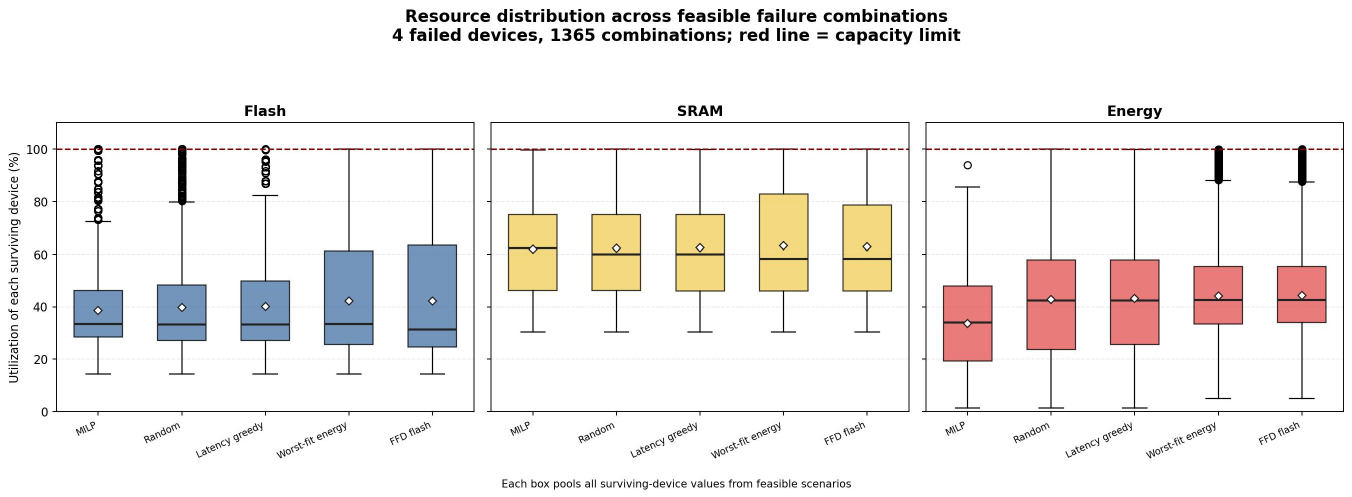}
    }

    \par\medskip

    \subfloat[\(F=5\): \(15504\) combinations
    \label{fig:resource_n20_f5}]{
        \includegraphics[width=0.47\textwidth]
        {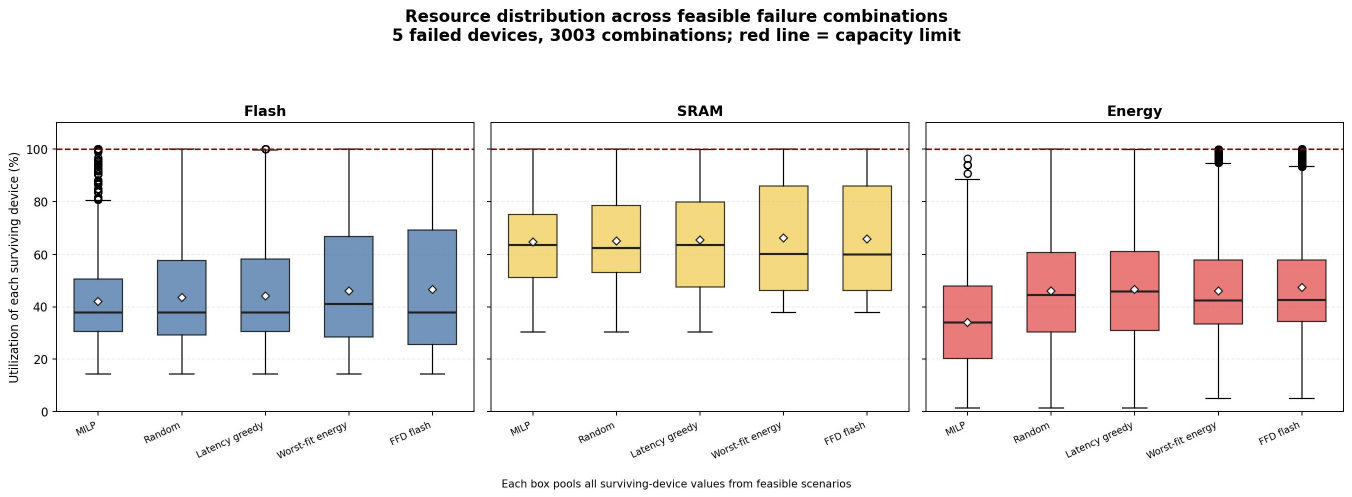}
    }

    \caption{Distribution of flash, SRAM, and energy utilization across
    the surviving devices for \(N=20\) initially active devices under
    (a) \(F=1\), (b) \(F=2\), (c) \(F=3\), (d) \(F=4\), and
    (e) \(F=5\) simultaneous device failures. For each failure count,
    all \(\binom{20}{F}\) distinct failure combinations were evaluated.
    Each boxplot pools the device-level utilization observations from
    the failure combinations for which the corresponding algorithm
    returned a feasible solution. The center line represents the median,
    the diamond represents the mean, the whiskers extend to
    \(1.5\times IQR\), and circles represent outliers. The red dashed
    line denotes the resource-capacity limit of \(100\%\). Lower values
    indicate greater remaining resource headroom.}

    \label{fig:resource_utilization_n20}
\end{figure*}
\fi


\section{Conclusion}\label{sec:conclusion}
Our approach restores \cgls{si} across heterogeneous
\cgls{iot} devices by deploying replicas through differential
\cgls{ota} updates, preserving the original partitioning
and surviving primary placements. The proposed \cgls{milp} optimizes replica placement while accounting for compatible communication protocols to minimize post-reconfiguration E2E inference latency under flash memory and runtime memory constraints. Simulations show lower average latency than four placement heuristics, with larger gains under reduced link rates in the 15-device network, while quantifying recovery-plan computation time. Future work will extend the optimization to OTA update
latency and device energy consumption.

\bibliographystyle{IEEEtran}
\bibliography{mybib}

\end{document}
\endinput